\documentclass[%
pre,
reprint,
superscriptaddress,
showpacs,
nofootinbib,
nobibnotes,
 amsmath,amssymb,
 aps,
]{revtex4-1}
\usepackage{geometry}                
\usepackage{graphicx}
\usepackage[sort&compress]{natbib}
\usepackage[caption=false]{subfig}
\usepackage{microtype}
\usepackage{dcolumn}
\usepackage{bm}
\usepackage{epstopdf}
\usepackage{amsthm}
\usepackage{xspace}
\usepackage{pifont} 
\usepackage{color}
\usepackage{listings}
\usepackage{rotating}
\usepackage{array,booktabs}
\newcommand{\gcc}{{\tt 403.gcc}\xspace}
\newcommand{\svd}{{\tt dgesdd}\xspace}
\newcommand{\svdone}{{\tt dgesdd}$_1$\xspace}
\newcommand{\svdtwo}{{\tt dgesdd$_2$}\xspace}
\newcommand{\svdthree}{{\tt dgesdd$_3$}\xspace}
\newcommand{\svdfour}{{\tt dgesdd$_4$}\xspace}
\newcommand{\svdfive}{{\tt dgesdd$_5$}\xspace}
\newcommand{\svdsix}{{\tt dgesdd$_6$}\xspace}
\newcommand{\arima}{{\tt auto.arima}\xspace}

\newcommand{\naive}{na\"ive}

\newcommand{\col}{{\tt col\_major}\xspace}

\newtheorem*{mydef}{Definition}

\begin{document}

\preprint{APS/123-QED}


\title{
Model-free quantification of time-series predictability
}


\author{Joshua Garland}
\email{joshua.garland@colorado.edu}
\affiliation{Department of Computer Science, University of Colorado at Boulder. Colorado, USA}

\author{Ryan James}
\email{ryan.james@colorado.edu}
\affiliation{Department of Computer Science, University of Colorado at Boulder. Colorado, USA}

\author{Elizabeth Bradley}
\email{lizb@colorado.edu}
\affiliation{Department of Computer Science, University of Colorado at Boulder. Colorado, USA}
\affiliation{Santa Fe Institute, New Mexico, USA}

\date{\today}

\begin{abstract}

This paper provides insight into when, why, and how forecast
strategies fail when they are applied to complicated time series.  We
conjecture that the inherent complexity of real-world time-series
data---which results from the dimension, nonlinearity, and
non-stationarity of the generating process, as well as from
measurement issues like noise, aggregation, and finite data
length---is both \emph{empirically} quantifiable and directly
correlated with predictability.  In particular, we argue that
\emph{redundancy} is an effective way to measure complexity and
predictive structure in an experimental time series and that
\emph{weighted permutation entropy} is an effective way to estimate
that redundancy.  To validate these conjectures, we study 120
different time-series data sets.  For each time series, we construct
predictions using a wide variety of forecast models, then compare the
accuracy of the predictions with the permutation entropy of that time
series.  We use the results to develop a model-free heuristic that can
help practitioners recognize when a particular prediction method is
not well matched to the task at hand: that is, when the time series
has more predictive structure than that method can capture and
exploit.

\vspace{0.1in}

\preprint{Santa Fe Institute Working Paper XX-XX-XXX}
\preprint{arxiv.org:XXXX.XXXX [nlin.CD]}

\end{abstract}

\maketitle

\section{Introduction}\label{sec:intro}
Complicated time-series data are ubiquitous in modern scientific
research.  The complexity of these data spans a wide range.  On the
low end of this spectrum are time series that exhibit perfect
predictive structure, i.e, signals whose future values can be
successfully predicted from past values.  Signals like this can be
viewed as the product of an underlying process that generates
information and/or transmits it from the past to the future in a
perfectly predictable fashion.  Constant or periodic signals, for
example, fall in this class.  On the opposite end of this spectrum are
signals that are what one could call \emph{fully complex}, where the
underlying generating process transmits no information at all from the
past to the future.  White noise processes fall in this class.  In
fully complex signals, knowledge of the past gives no insight into the
future, regardless of what model one chooses to use. Signals in the
midrange of this spectrum, e.g., deterministic chaos, pose interesting
challenges from a modeling perspective.  In these signals, enough
information is being transmitted from the past to the future that an
\emph{ideal} model---one that captures the generating process---can
forecast the future behavior of the observed system with high
accuracy.


This leads naturally to an important and challenging question: given a
noisy real-valued time series from an unknown system, does there exist
a forecast model that can leverage the information (if any) that is
being transmitted forward in time by the underlying generating
process?  A first step in answering this question is to reliably
quantify where on the complexity spectrum a given time series falls; a
second step is to determine how complexity and predictability are
related in these kinds of data sets.  With these answers in hand, one
can develop a practical strategy for assessing appropriateness of
forecast methods for a given time series.  If the forecast produced by
a particular method is poor, for instance, but the time series
contains a significant amount of predictive structure, one can
reasonably conclude that that method is inadequate to the task and
that one should seek another method.  The goal of this paper is to
develop effective heuristics to put that strategy into practice.


The information in an observation can be partitioned into two pieces:
redundancy and entropy generation~\cite{crutchfield2003}.
\label{page:redundancy}
Our approach exploits this decomposition in order to assess how much
predictive structure is present in a signal---i.e., where it falls on
the complexity spectrum mentioned above.  We define \emph{complexity}
as a particular approximation of Kolmogorov-Sinai
entropy~\cite{lind95}.  That is, we view a random-walk time series
(which exhibits high entropy) as purely complex, whereas a low-entropy
periodic signal is on the low end of the complexity spectrum.  This
differs from the notion of complexity used by e.g. \cite{Shalizi2008},
which would consider a time series without any statistical
regularities to be non-complex.  We argue that an extension of
\emph{permutation entropy}~\cite{bandt2002per}---a method for
approximating the entropy through ordinal analysis---is an effective
way to assess the complexity of a given time series.  Permutation
entropy is ideal for our purposes because it works with real-valued
data and is known to converge to the true entropy value. Other
existing techniques either require specific knowledge of the
generating process or produce biased values of the
entropy~\cite{bollt2001}.

We focus on real-valued, scalar, time-series data from physical experiments.
%
%
We do not assume any knowledge of the generating process or its
properties: whether it is linear, nonlinear, deterministic,
stochastic, etc.  To explore the relationship between complexity,
predictive structure, and actual predictability, we generate forecasts
for a variety of experimental time-series datasets using four
different prediction methods, then compare the accuracy of those
predictions to the permutation entropy of the associated signals.
This results in two primary findings:
\begin{enumerate}

\item The permutation entropy of a noisy real-valued time series from
  an unknown system is correlated with the accuracy of an appropriate
  predictor.

\item The relationship between permutation entropy and prediction
  accuracy is a useful empirical heuristic for identifying mismatches
  between prediction models and time-series data.

\end{enumerate}
There has, of course, been a great deal of good work on different ways
to measure the complexity of data, and previous explorations have
confirmed repeatedly that complexity is a challenge to prediction.  It
is well known that the way information is generated and processed
internally by a system plays a critical role in the success of
different forecasting methods---and in the choice of which method is
appropriate for a given time series.  This constellation of issues has
not been properly explored, however, in the context of noisy, poorly
sampled, real-world data from unknown systems.  That exploration, and
the development of strategies for putting its results into effective
practice, is the primary contribution of this work.  The empirical
results in the later sections of this paper not only elucidate the
relationship between complexity and predictability.  The strategy that
we derive from these results can also aid practitioners in choosing an
appropriate prediction model for a given real-world noisy time series
from an unknown system---a challenging task for which little guidance
is currently available.

The rest of the paper is organized as follows.
Section~\ref{sec:related} discusses previous results on generating
partitions, local modeling, and error distribution analysis, and
situates this work in that context. Section~\ref{sec:methods} covers
the experimental setup and methods used to collect the time-series
data. Section~\ref{sec:model} describes the prediction models used in
this study.  Section~\ref{sec:meaComplex} reviews permutation entropy,
the technique that we use to measure complexity.  In
Section~\ref{sec:results}, we estimate the complexity of each
empirical time series and compare that complexity to the accuracy of
predictions produced by the methods of Section~\ref{sec:model},
operating on that time series.  In Section~\ref{sec:conc}, we discuss
these results and their implications, and consider future areas of
research.

\section{Related Work }\label{sec:related}

Modeling time-series data for the purposes of prediction dates back at
least to Yule's 1927 invention of autoregression~\cite{Yule27}.  Since
then, hundreds, if not thousands, of strategies have been developed
for a wide variety of prediction tasks.  The purpose of this paper is
not to add a new weapon to this arsenal, nor to do any sort of
theoretical assessment or comparison of existing methods.  Our goals
are focused more on the \emph{practice} of prediction: \emph{(i)} to
empirically quantify the predictive structure that is present in a
real-valued scalar time series and \emph{(ii)} to explore how the
performance of prediction methods is related to that inherent
complexity.  It would, of course, be neither practical nor interesting
to report results for every existing forecast method; instead, we
choose a representative set, as described in Section~\ref{sec:model}.

Quantifying predictability, which is sometimes called ``predicting
predictability,'' is not a new problem.  Most of the corresponding
solutions fall into two categories that we call model-based error
analysis and model-free information analysis.
%
%
%
The first class focuses on errors produced by a fixed forecasting
schema.  This analysis can proceed locally or globally.  The local
version approximates error distributions for different regions of a
time-series model using local ensemble in-sample
forecasting\footnote{The terms ``in sample'' and ``out of sample'' are
  used in different ways in the forecasting community.  Here, we
  distinguish those terms by the part of the time series that is the
  focus of the prediction: the observed data for the former and the
  unknown future for the latter.  In-sample forecasts---comparisons of
  predictions generated from \emph{part} of the observed time
  series---are useful for assessing model error and prediction
  horizons, among other things.}.
These distributions are then used as estimates of out-of-sample
forecast errors in those regions.  For example, Smith \emph{et al.}
make in-sample forecasts using ensembles around selected points in
order to predict the local predictability of that time
series~\cite{Smith199250}.  This approach can be used to show that
different portions of a time series exhibit varying levels of
local predictive uncertainty.  We expand on this idea later in this
paper with a time series that exhibits interesting regime shifts.

Local model-based error analysis works quite well, but it only
approximates the \emph{local} predictive uncertainty \emph{in relation
  to a fixed model}.  It cannot quantify the inherent predictability
of a time series and thus cannot be used to draw conclusions about
predictive structure that may be usable by other forecast methods.
%
%
Global model-based error analysis moves in this direction.  It uses
out-of-sample error distributions, computed \emph{post facto} from a
class of models, to determine which of those models was best.  After
building an autoregressive model, for example, it is common to
calculate forecast errors and verify that they are normally
distributed.  If they are not, that suggests that there is structure
in the time series that the model-building process was unable to
recognize, capture, and exploit.  The problem with this approach is
lack of generality.
\label{page:normal-errors}
Normally distributed errors indicate that a model has captured the
structure in the data insofar as is possible, \emph{given the
  formulation of that particular model} (viz., the best possible
linear fit to a nonlinear dataset).  This gives no indication as to
whether another modeling strategy might do better.

A practice known as deterministic vs. stochastic
modeling~\cite{weigend93, Casdagli92dvsplots} bridges the gap
between local and global approaches to model-based error analysis.
The basic idea is to construct a series of local linear fits,
beginning with a few points and working up to a global linear fit that
includes all known points, and then analyze how the average
out-of-sample forecast error changes as a function of number of points
in the fit. The shape of such a ``DVS" graph indicates the amounts of
determinism and stochasticity present in a time series.

The model-based error analysis methods described in the previous three
paragraphs are based on specific assumptions about the underlying
generating process and knowledge about what will happen to the error
if those assumptions hold or fail.  Model-\emph{free} information
analysis moves away from those restrictions.  Our approach falls into
this class: we wish to measure the inherent complexity of an empirical
time series, then study the correlation of that complexity with the
predictive accuracy of forecasts made using a number of different
methods.

We build on the notion of \emph{redundancy} that was introduced on
page~\pageref{page:redundancy}, which formally quantifies how
information propagates forward through a time series:
i.e., the mutual information between the past $n$ observations and the
current one.
%
The redundancy of i.i.d. random processes, for instance, is zero,
since all observations in such a process are independent of one
another.  On the other hand, deterministic systems---including chaotic
ones---have high redundancy that is maximal in the infinite limit, and
thus they can be perfectly predicted if observed for long
enough~\cite{weigend93}.  In practice, it is quite difficult to
estimate the redundancy of an arbitrary, real-valued time series.
Doing so requires knowing either the Kolmogorov-Sinai entropy or the
values of all positive Lyapunov exponents of the system.  Both of
these calculations are difficult, the latter particularly so if the
data are very noisy or the generating system is stochastic.

Using entropy and redundancy to quantify the inherent predictability
of a time series is not a new idea.  Past methods for this, however,
(e.g.,~\cite{Shannon1951, mantegna1994linguistic}) have hinged on
knowledge of the \emph{generating partition} of the underlying
process, which lets one transform real-valued observations into
symbols in a way that preserves the underlying dynamics~\cite{lind95}.
Using a projection that is not a generating partition---e.g., simply
binning the data---can introduce spurious complexity into the
resulting symbolic sequence and thus misrepresent the entropy of the
underlying system~\cite{bollt2001}.  Generating partitions are
luxuries that are rarely, if ever, afforded to an analyst, since one
needs to know the underlying dynamics in order to construct one.  And
even if the dynamics are known, these partitions are difficult to
compute and often have fractal boundaries~\cite{eisele1999}.

We sidestep these issues by using a variant of the \emph{permutation
  entropy} of Bandt and Pompe~\cite{bandt2002per} to estimate the
value of the Kolmogorov-Sinai entropy of a real-valued time
series---and thus the redundancy in that data, which our results
confirm to be an effective proxy for predictability.  This differs
from existing approaches in a number of ways.  It does not rely on
generating partitions---and thus does not introduce bias into the
results if one does not know the dynamics or cannot compute the
partition.  Permutation entropy makes no assumptions about, and
requires no knowledge of, the underlying generating process: linear,
nonlinear, the Lyapunov spectrum, etc.  These features make our
approach applicable to noisy real-valued time series from all classes
of systems, deterministic and stochastic.

There has been prior work under a very similar title to
ours~\cite{haven2005}, but there are only superficial similarities
between the two research projects. Haven \emph{et al.} utilize the
relative entropy to quantify the difference in predictability between
two distributions: one evolved from a small ensemble of past states
using the known dynamical system, and the other the observed
distribution. Our work quantifies the predictability of a single
observed time series using weighted permutation entropy and makes no
assumptions about the generating process.

More closely related is the work of Boffetta \emph{et
  al.}~\cite{boffetta02}, who investigate the scaling behavior of
finite-size Lyapunov exponents (FSLE) and $\epsilon$-entropy for a
wide variety of deterministic systems with known dynamics and additive
noise.  While the scaling of these measures acts as a general proxy
for predictability bounds, this approach differs from our work in a
number of fundamental ways.  First, \cite{boffetta02} is a theoretical
study that does not involve any actual predictions.  We focus on
real-world time-series data, where one does not necessarily have the
ability to perturb or otherwise interact with the system of interest,
nor can one obtain or manufacture the (possibly large) number of
points that might be needed to estimate the $\epsilon$-entropy for
small $\epsilon$.  Second, we do not require \emph{a priori} knowledge
about the noise and its interaction with the system.  Third, we tie
information---in the form of the weighted permutation
entropy---directly to prediction error via calculated values of a
specific error metric.  Though FSLE and $\epsilon$-entropy allow for
the comparison of predictability between systems, they do not directly
provide an estimate of prediction error.  Finally, our approach also
holds for stochastic systems, where neither the FLSEs nor their
relationship to predictability is well defined.
\section{Experimental Methods}\label{sec:methods}

\begin{figure}[htp]
    \centering
    \includegraphics[width=\columnwidth]{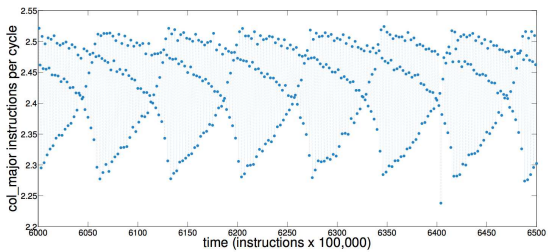}
    \caption{A short segment of a computer performance trace: the
      instructions per CPU clock cycle (IPC) during the execution of
      \col, a simple program that repeatedly initializes a matrix in
      column-major order.  Each point is the average IPC in a 100,000
      instruction period.}
   \label{fig:col-ipc}
  \end{figure}

For the purposes of this study, we required a broad array of
time-series datasets from across the complexity spectrum.  We chose to
study sensor data from a computer-performance experiment.  While this
is not a common laboratory experiment, it is a highly appropriate
choice here.  Computers are extremely complicated systems and their
dynamics is surprisingly rich.
The processor and memory loads during the execution of even a very
simple program can exhibit dynamical chaos, for
instance~\cite{mytkowicz09}.  Figure~\ref{fig:col-ipc} shows an
example: a short segment of a performance trace of a four-line C
program that repeatedly initializes the upper triangle of a matrix in
column-major order.
A small change in the code can cause the dynamics to bifurcate to a
periodic regime.  
%
By running different programs on the same computer, then, we can
produce traces that span the whole range of the complexity spectrum,
from completely predictable to completely unstructured---which makes
this an ideal testbed for this study\footnote{Predicting the
  \emph{state} of a computer, of course, would amount to solving the
  halting problem.  What we are doing here is predicting computer
  \emph{performance}, which does not violate the Rice-Shapiro
  theorem~\cite{hopcroft2007}.}.

The time-series data sets for these experiments were collected on an
Intel Core\textsuperscript{\textregistered} i7-2600-based machine.
%
%
We gathered performance traces during the execution of three different
programs---the simple \col loop whose performance is depicted in
Figure~\ref{fig:col-ipc} and two more-complex programs: one from the
SPEC 2006CPU benchmark suite (\gcc), and one from the LAPACK linear
algebra package (\svd).  In all of these experiments, the scalar
observation $x_i$ was a measurement of the processor performance at
time $i$ during the execution of each program.
%
%
%
For statistical validation, we collected 15 performance traces from
each of the three programs.  For an in-depth description of the
experimental setup used to gather these data,
please see
\cite{josh-ida2011,josh-ida2013,zach-IDA10,mytkowicz09,todd-phd}.

\begin{figure}[t]
  \centering
    \includegraphics[width=\columnwidth]{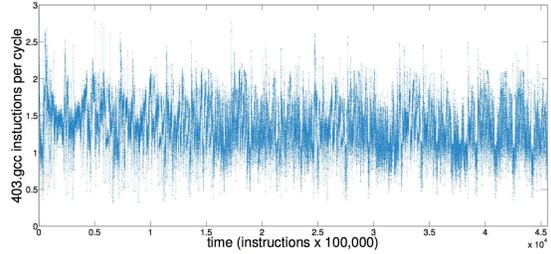}
    \caption{An IPC trace during the execution of \gcc.}
    \label{fig:gcc-ts}
  \end{figure}

The SPEC CPU2006 benchmark suite \cite{spec} is a collection of
complicated programs that are used in the computer-science community
to assess and compare the performance of different computers.  \gcc is
a member of that suite.
%
%
It is a \emph{compiler}: a program that translates code written in a
high-level language 
into a lower-level format that can be executed by the processor chip.
Its behavior is far more complicated than that of \col, as is clear
from Figure~\ref{fig:gcc-ts}.
Unlike \col, where the processor utilization is quite structured,
\gcc's performance appears almost random.

\begin{figure}[t]
    \centering
    \includegraphics[width=\columnwidth]{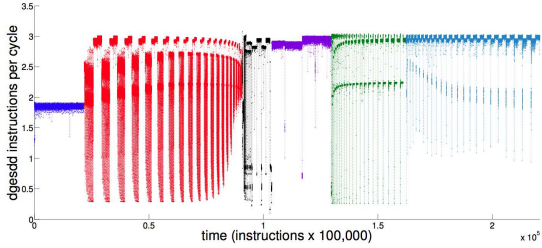}
    \caption{An IPC trace during the execution of \gcc.  The colors
      identify the different \emph{segments} of the signal that are
      discussed in the text.}
    \label{fig:svd-ts-colored}
  \end{figure}

\svd is a Fortran program from the LAPACK linear algebra package
\cite{lapack}.  It calculates the singular value decomposition of a
rectangular $M$ by $N$ matrix with real-valued entries.  For our
experiments, we chose $M=750$ and $N=1000$ and generated the matrix
entries randomly.
%
%
The behavior of this program as it computes the singular values of
this matrix is very interesting, as is clearly visible in Figure
\ref{fig:svd-ts-colored}.  
As the code moves though its different phases---diagonalizing the
matrix, computing its transpose, multiplying, etc.---the processor
utilization patterns change quite radically.  For the first
$\sim$21,000 measurements (21,000 $\times$ 100,000 instructions),
roughly 1.8 instructions are executed per cycle, on the average, by
the eight processing units on this chip.  After that, the IPC moves
through a number of different oscillatory regimes, which we have
color-coded in the figure in order to make textual cross-references
easy to track.

The wide range of behaviors in Figure~\ref{fig:svd-ts-colored}
provides a distinct advantage, for the purposes of this paper, in that
a number of different generating processes---with a wide range of
complexities---are at work in different phases of a single time
series.  The \col and \gcc traces in Figures~\ref{fig:col-ipc} and
\ref{fig:gcc-ts} appear to be far more consistent over time---probably
the result of a single generating process with consistent complexity.
\svd, in contrast, has multiple regimes, each the result of different
generating processes.  To take advantage of this, we split the signal
into six different segments, thereby obtaining an array of examples
for the analyses in the following sections.  For notational
convenience, we refer to these 90 time-series data sets\footnote{15
  runs, each with six regimes} as {\tt dgesdd$_i$}, with $i \in
\{1\dots6\}$ where $i$ corresponds to one of the six segments of the
signal, ordered from left to right.  These segments, which were
determined visually, are shown in different colors in
Figure~\ref{fig:svd-ts-colored}.  Visual decomposition is subjective,
of course, particularly since the regimes exhibit some fractal
structure.  Thus, it may be the case that more than one generating
process is at work in each of our segments.  This is a factor in the
discussion of Section~\ref{sec:results}.


\section{Modeling }\label{sec:model}
%
%
%
%

In this section, we describe the four different forecasting methods
used in this study, as well as the error metric used to evaluate their
predictive accuracy.  These methods include:
\begin{itemize}
\item The \emph{random-walk} method, which uses the previous value in
  the observed signal as the forecast,

\item The \emph{\naive} method, which uses the mean of the
  observed signal as the forecast,

\item The \emph{ARIMA} (auto-regressive integrated moving average)
  method, a common linear forecast strategy, instatiated via the
  \emph{\arima} procedure \cite{autoARIMA}, and

\item The \emph{LMA} (Lorenz method of analogues) method, which uses a
  near-neighbor forecast strategy on a dynamical reconstruction of the
  signal.
\end{itemize}
ARIMA models are based on standard linear techniques.  LMA is designed
to capture and exploit the deterministic structure of a signal from a
nonlinear dynamical system.  The \naive ~and random-walk methods,
somewhat surprisingly, often outperform these more-sophisticated
prediction strategies in the case of highly complex signals, as
discussed below.


\subsection{Two Simple Prediction Strategies}
\label{sec:simple}

A random-walk predictor simply uses the last observed measurement as
the forecast: that is, the predicted value $p_i$ at time $i$ is
calculated using the following relation: $$p_i = x_{i-1}$$ The
prediction strategy that we refer to using the term ``\naive''
averages the prior observations to generate the forecast: $$p_i =
\sum_{j=1}^{i-1}\frac{x_j}{i-1}$$ While both of these methods are
simplistic, they are not without merit.  For a time series near the
high end of the complexity spectrum---i.e., one that possesses very
little predictive structure---these two methods can actually be the
best choice.  In forecasting currency exchange rates, for instance,
sophisticated econometrics-based prediction models fail to
consistently outperform the random-walk method~\cite{rwMeese,rwCCE}.
These signals are constantly changing, noisy, and possess very little
predictive structure, but their variations are not---on the
average---very large, so the random-walk method's strategy of simply
guessing the last known value is not a bad choice.  If a signal has a
unimodal distribution with low variance, the \naive ~prediction
strategy will perform quite well---even if the signal is highly
complex---simply because the mean is a good approximation of the
future behavior.  Moreover, the \naive ~prediction strategy's temporal
average effects a low-pass filtering operation, which can  mitigate the
complexity in signals with very little predictive structure.

Both of these methods have significant weaknesses, however.  Because
they do not model the temporal patterns in the data, or even the
distribution of its values, they cannot track changes in that
structure.  This causes them to fail in a number of important
situations.  Random-walk strategies are a particularly bad choice for
time series that change significantly at every time step.  In the
worst case---a large-amplitude square wave whose period is equivalent
to twice the sample time---a random-walk prediction would be exactly
180 degrees out of phase with the true continuation.  The \naive
~method would be a better choice in this situation, since it would
always split the difference.  It would, however, perform poorly when a
signal has a number of long-lived regimes that have significantly
different means.  In this situation, the inertia of the \naive
~method's accumulating mean is a liability and the agility of the
random-walk method is an advantage, since it can respond quickly to
regime shifts.

Of course, methods that could capture and exploit the geometry of the
data and/or its temporal patterns would be far more effective in the
situations described in the previous paragraph.  The \arima and LMA
methods introduced in Sections~\ref{sec:arima} and~\ref{sec:lma} are
designed to do exactly that.
However, if a signal contains little predictive structure, forecast
strategies like ARIMA and LMA have nothing to work with and thus will
often be outperformed by the two simple strategies described in this
section.  This effect is explored further in Sections~\ref{sec:accuracy}
and~\ref{sec:results}.

\subsection{A Regression-Based Prediction Strategy}
\label{sec:arima}

A simple and yet powerful way to capture and exploit the structure of
data is to fit a hyperplane to the known points and then use it to
make predictions.  The roots of this approach date back to the
original autoregressive schema~\cite{weigend93}, which forecasts the
next time step through a weighted average of past observations: $$p_i
= \sum_{j=1}^{i-1} a_j x_j$$ The weighting coefficients $a_j$ are
generally computed using either an ordinary least squares approach, or
with the method of moments using the Yule-Walker equations.  To
account for noise in the data, one can add a so-called ``moving
average'' term to the model; to remove nonstationarities, one can
detrend the data using a differencing operation.  A strategy that
incorporates all three of these features is called a \emph{nonseasonal
  ARIMA model}.  If evidence of periodic structure is present in the
data, a \emph{seasonal ARIMA model}, which adds a sampling operation
that filters out periodicities, can be a good choice.

There is a vast amount of theory and literature regarding the
construction and use of models of this type; we refer the reader to
\cite{davislinearts} for an in-depth exploration.  For the purposes of
this paper, where the goal is to explore the relationship between
predictability and complexity across a broad array of forecast
strategies, seasonal ARIMA models are a good exemplar of the class of
linear predictors.  Fitting such a model to a dataset involves
choosing values for the various free parameters in the autoregressive,
detrending, moving average, and filtering terms.  We employ the
automated fitting techniques described in~\cite{autoARIMA} to
accomplish this, producing what we will call an ``\arima model'' in
the rest of this paper.  This procedure uses sophisticated
methods---KPSS unit-root tests~\cite{KPSSunit}, a customization of the
Canova-Hansen test~\cite{Canova1995}, and the Akaike information
criterion~\cite{akaike1974}, conditioned on the maximum likelihood of
the model fitted to the detrended data---to select good values for the
free parameters of the ARIMA model.

ARIMA forecasting is a common and time-tested procedure.  Its
adjustments for seasonality, nonstationarity, and noise make it an
appropriate choice for short-term predictions of time-series data
generated by a wide range of processes.  If information is being
generated and/or transmitted in a nonlinear way, however, a global
linear fit is inappropriate and ARIMA forecasts can be inaccurate.
Another weakness of this method is prediction horizon: an ARIMA
forecast is guaranteed to converge (to the mean, to a constant value,
or to a linear trend) after some number of predictions, depending on
model order.  To sidestep this issue, we build forecasts in a stepwise
fashion: i.e., fit the \arima model to the existing data, use that
model to perform a one-step prediction, rebuild the \arima model using
the latest observations, and iterate until the desired prediction
horizon is reached.  For consistency, we take the same approach with
the other three models in this study as well, even though doing so
amounts to artificially hobbling LMA.

\subsection{A Nonlinear Prediction Strategy}
\label{sec:lma}

When the temporal progressions in a time series are produced by a
deterministic nonlinear process, one can use a technique called
delay-coordinate embedding
%
%
to model the structure of the information generation and transmission
occurring in the underlying process, then use that reconstruction to
generate forecasts.  This section discusses the theory and
implementation of a prediction strategy that is based on this idea.

Delay-coordinate embedding~\cite{packard80,Sauer:1991lr,Takens:1981uq}
allows one to reconstruct a dynamical system's full state-space
dynamics from a scalar time-series measurement---provided that some
conditions hold regarding those data.  Specifically, if the underlying
dynamics and the measurement function---the mapping from the unknown
state vector $\vec{X}$ to the observed value $x_i$---are both smooth
and generic, Takens~\cite{Takens:1981uq} formally proves that the
delay-coordinate map
\[
F(\tau,m)(\vec{X}) = ([x_{i} ~ x_{i+\tau} ~ \dots ~x_{i+m\tau}])
\]
from a $d$-dimensional smooth compact manifold $M$ to
$\mathbb{R}^{2d+1}$ is a diffeomorphism on $M$: in other words, that
the reconstructed dynamics and the true (hidden) dynamics have the
same topology.  This is an extremely powerful result; among other
things, it means that one can model the full system dynamics, up to
diffeomorphism, without measuring---or even knowing---anything about
the state variables.

\begin{figure}[!ht]
    \subfloat[\col\label{fig:colEmbedding}]{%
      \includegraphics[width=\columnwidth]{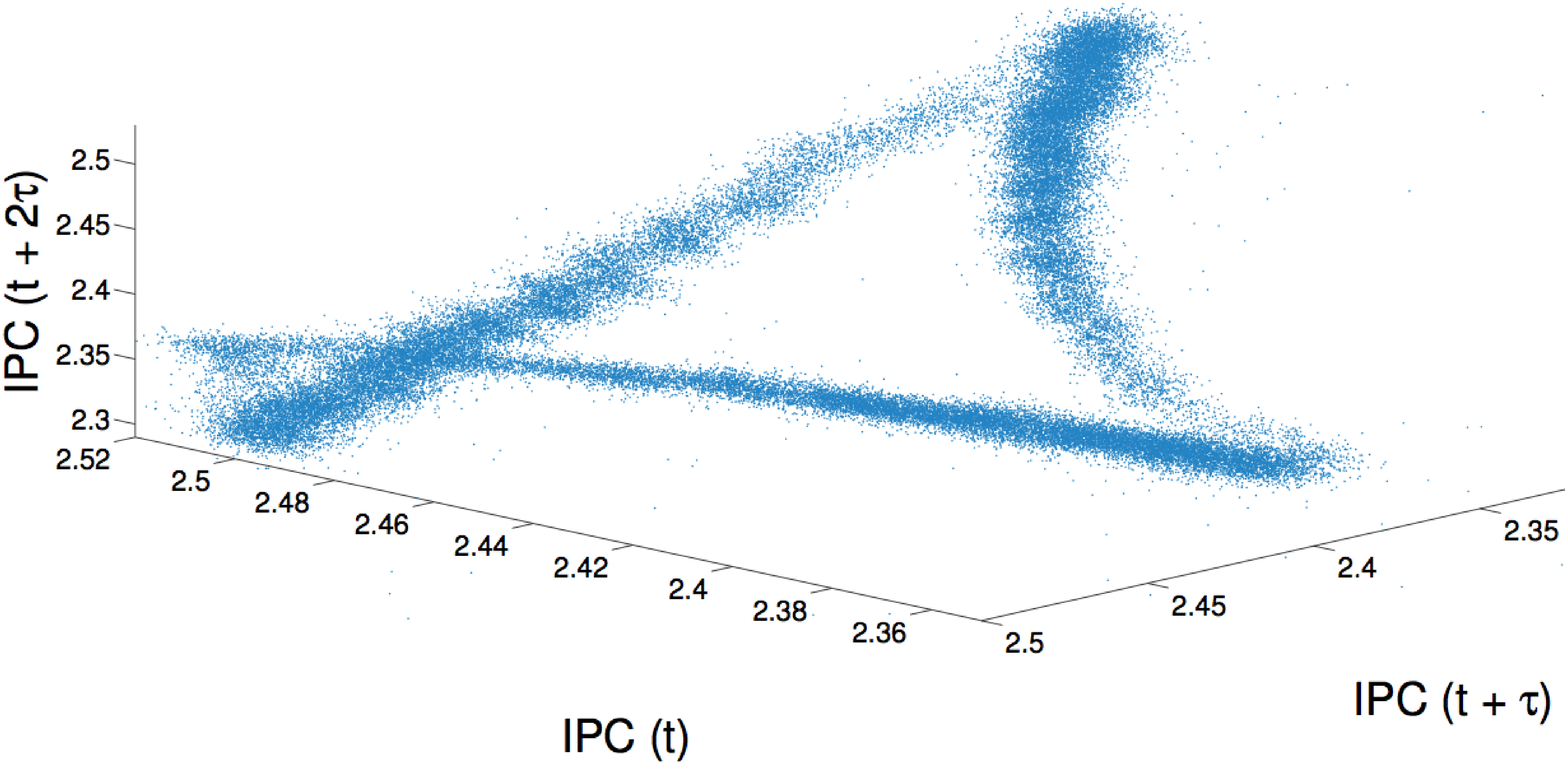}
    }
    \hfill
    \subfloat[\gcc\label{fig:gccEmbedding}]{%
      \includegraphics[width=\columnwidth]{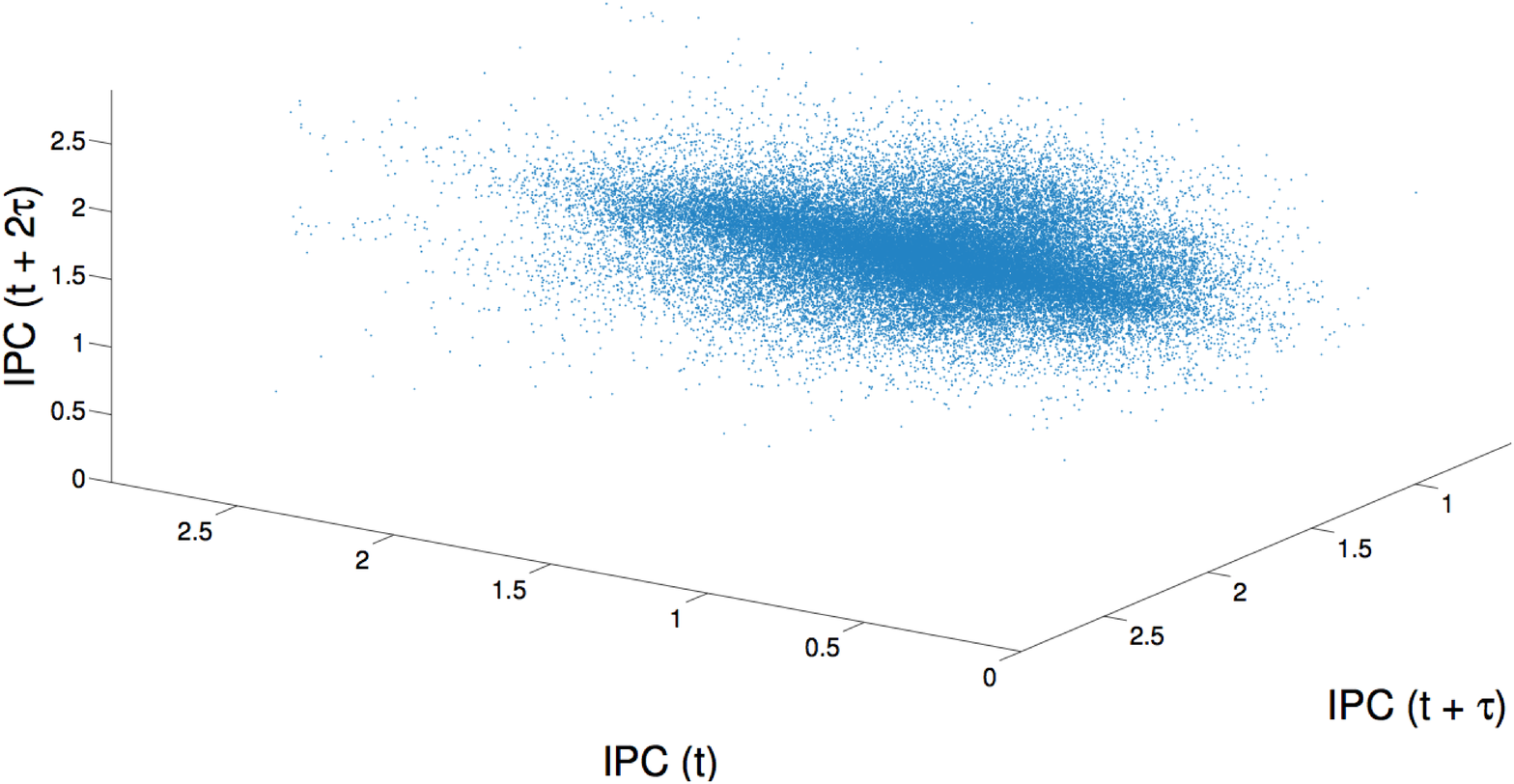}
    }
        \hfill
    \subfloat[\svdfive \label{fig:svdfiveEmbedding}]{%
      \includegraphics[width=\columnwidth]{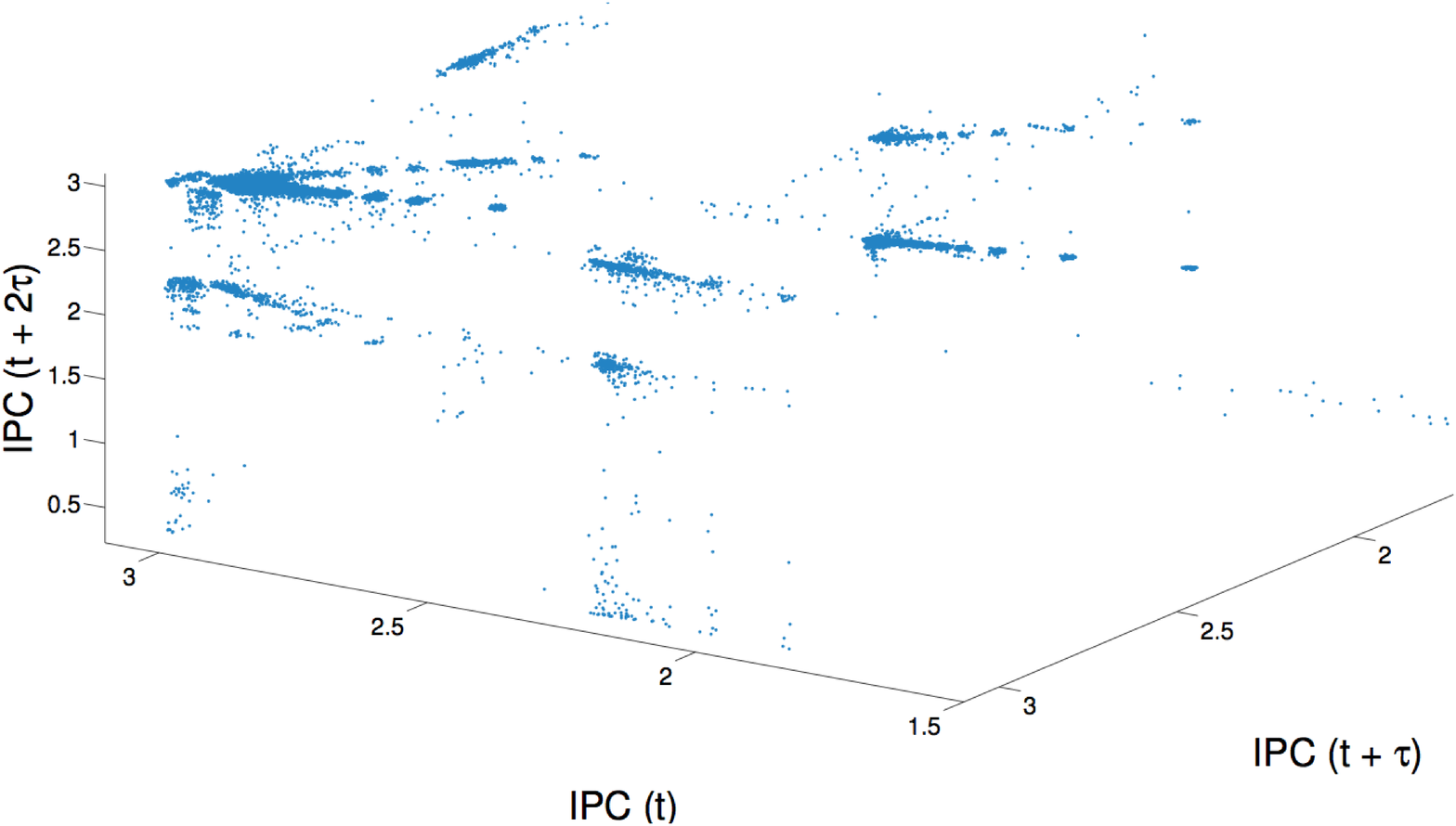}
    }
    \caption{3D projections of delay-coordinate embeddings of the
       traces from (a) Figure~\ref{fig:col-ipc} (b)
       Figure~\ref{fig:gcc-ts} and (c) the fifth (green) segment of
       Figure~\ref{fig:svd-ts-colored}.}
    \label{fig:embedding}
  \end{figure}

The first step in the delay-coordinate embedding process is to
estimate values for the two free parameters in the map: the delay
$\tau$ and the dimension $m$.  We follow standard procedures for this,
choosing the first minimum in the time-delayed mutual information as
an estimate of $\tau$~\cite{fraser-swinney} and using the
false-near(est)-neighbor method of~\cite{KBA92} to estimate $m$.  Some
example plots of data from
Figures~\ref{fig:col-ipc}-\ref{fig:svd-ts-colored}, embedded following
this procedure, are shown in Figure~\ref{fig:embedding}.
%
%
%
%

Geometric structure in these kinds of plots is an indication of
structure in the information generation/transmission process that
produced the time series.  The dynamical systems community has
developed a number of methods that leverage this structure to generate
predictions (e.g.,~\cite{weigend-book,casdagli-eubank92,Smith199250}).
One of the most straightforward of these is the \emph{Lorenz method of
  analogues} (LMA), which is essentially nearest-neighbor prediction
in the embedded\footnote{Lorenz's original formulation used the full
  system state space;
%
%
this method was first extended to embedded dynamics by
Pikovsky~\cite{pikovsky86-sov}, but is also related to the prediction
work of Sugihara \& May~\cite{sugihara90}}
space~\cite{lorenz-analogues}.  Even this simple algorithm---which
builds predictions by finding the nearest neighbor in the embedded
space of the given point, then taking that neighbor's path as the
prediction---provides remarkably accurate forecasts when the
generating process is a deterministic dynamical system.

Since LMA does not rest on an assumption of linearity (as ARIMA models
do), it can handle both linear and nonlinear processes.  If the
underlying generating process is nondeterministic, however, it can
perform poorly.  Figure~\ref{fig:embedding}(b), for instance, appears
to contain little structure, so one might not expect LMA to work well
on this signal.  More structure appears to be present in
Figure~\ref{fig:embedding}(c), but this reconstruction also appears to
contain some noise.  The question as to how much structure is present
in a reconstruction, and how much of that structure can be captured
and used by LMA, is apropos of the central question treated in this
paper.  It may be that \gcc has some redundancy that LMA cannot
exploit, or that the structure in \svdfive is effectively obfuscated,
from the standpoint of the LMA method, by noise.  

Without any knowledge of the generating process, answers to these
questions can only be derived from the data, with all of the attendant
problems (noise, sampling issues, and so on).  By quantifying the
balance between redundancy, predictive structure, and entropy for
these real-valued time series---as shown in
Section~\ref{sec:results}---we can begin to answer these questions in
an effective and practical manner.


\subsection{Assessing Prediction Accuracy}
\label{sec:accuracy}

To study the relationship between predictability and complexity, we
use the four methods outlined above to generate predictions of all 120
traces described in Section~\ref{sec:methods}, then calculate the
error of the predictions with respect to the true continuations.
Specifically, we split each time series into two pieces: the first
90\%, referred to here as the ``initial training" signal and denoted
$\{x_i\}_{i=1}^{n}$, and the last 10\%, known as the ``test" signal
$\{c_j\}_{j=n+1}^{k+n+1}$.  The initial training signal is used to
build the model, following the procedures described in the previous
section; that model is used to generate a prediction of the value of
$x_{n+1}$, which is then compared to the true continuation, $c_{n+1}$.
The model is then rebuilt using $\{x_i\}_{i=1}^{n+1}$ and the process
repeats $k$ times, out to the end of the observed time series.  This
``one step prediction'' process is not technically necessary in the
LMA method, whose ability to generate accurate predictions is limited
only by the positive Lyapunov exponents of the system.  However, the
performance of the other three methods used here will degrade severely
if the associated models are not periodically rebuilt.  In order to
make the comparison fair, we used an iterative one-step prediction
schema \emph{for all four methods}.  This has the slightly confusing
effect of causing the ``test'' signal to be used both to assess the
accuracy of each model and for periodic refitting.

\begin{figure*}
  \renewcommand{\arraystretch}{1.25}
  \begin{tabular}{m{0.1\columnwidth}|CCC}
    & {\large \col} & {\large \gcc} & {\large \svdfive} \\
    & & & \\
            \hline \\
    \rotatebox{90}{\large{random walk}} &
    \includegraphics[width=0.6\columnwidth]{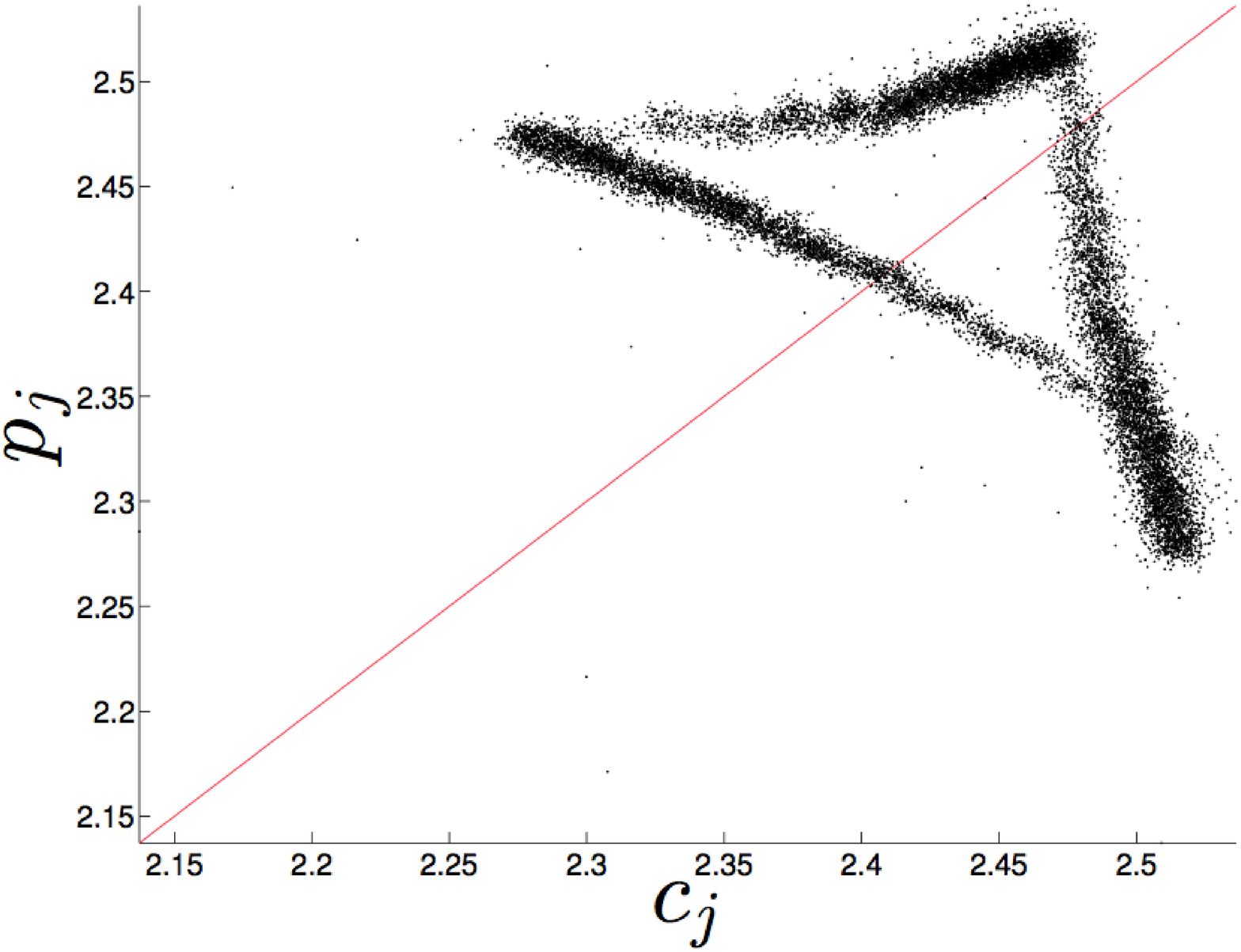} &
    \includegraphics[width=0.6\columnwidth]{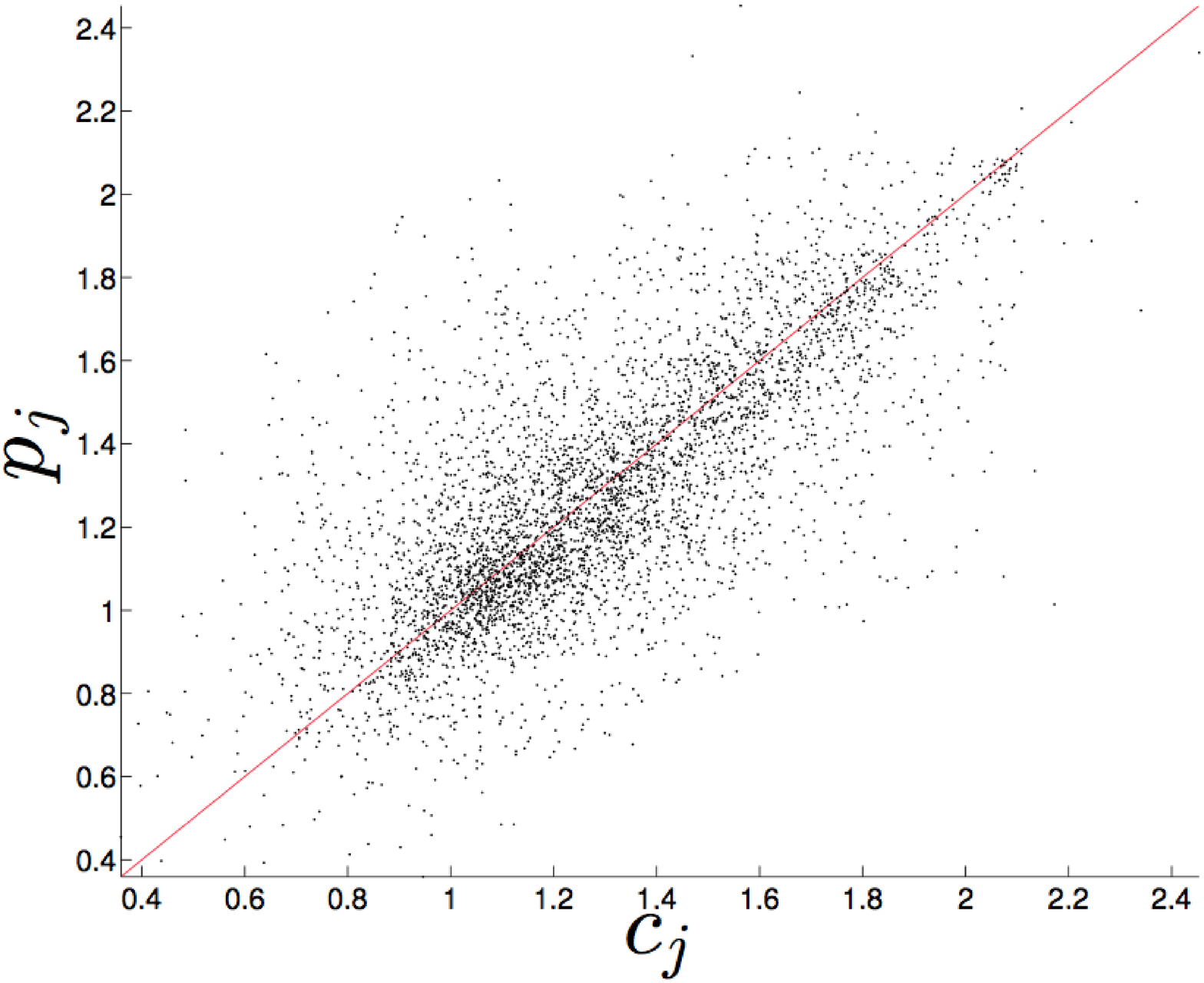} &
    \includegraphics[width=0.6\columnwidth]{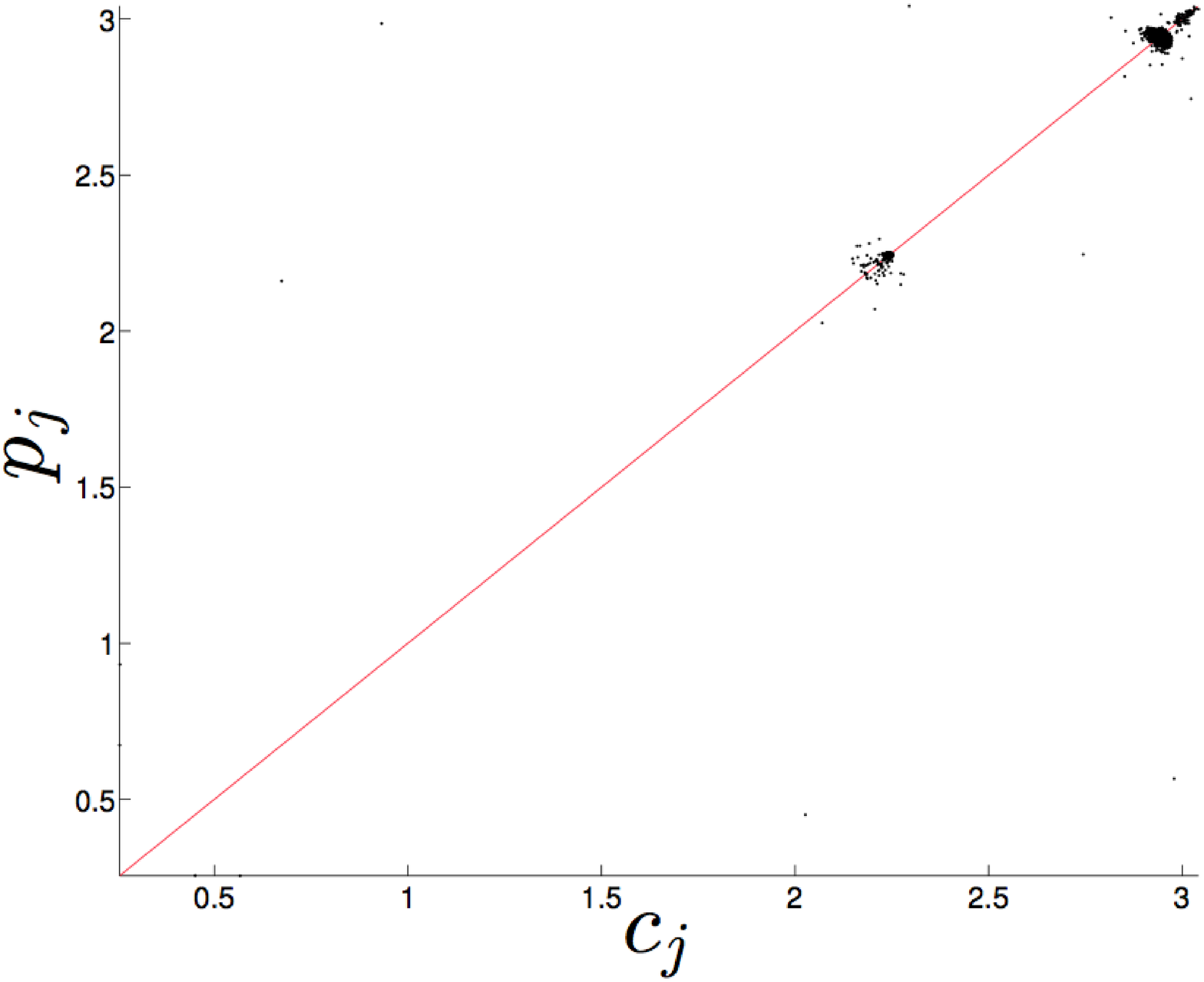} \\
    \rotatebox{90}{\large \naive} &
    \includegraphics[width=0.6\columnwidth]{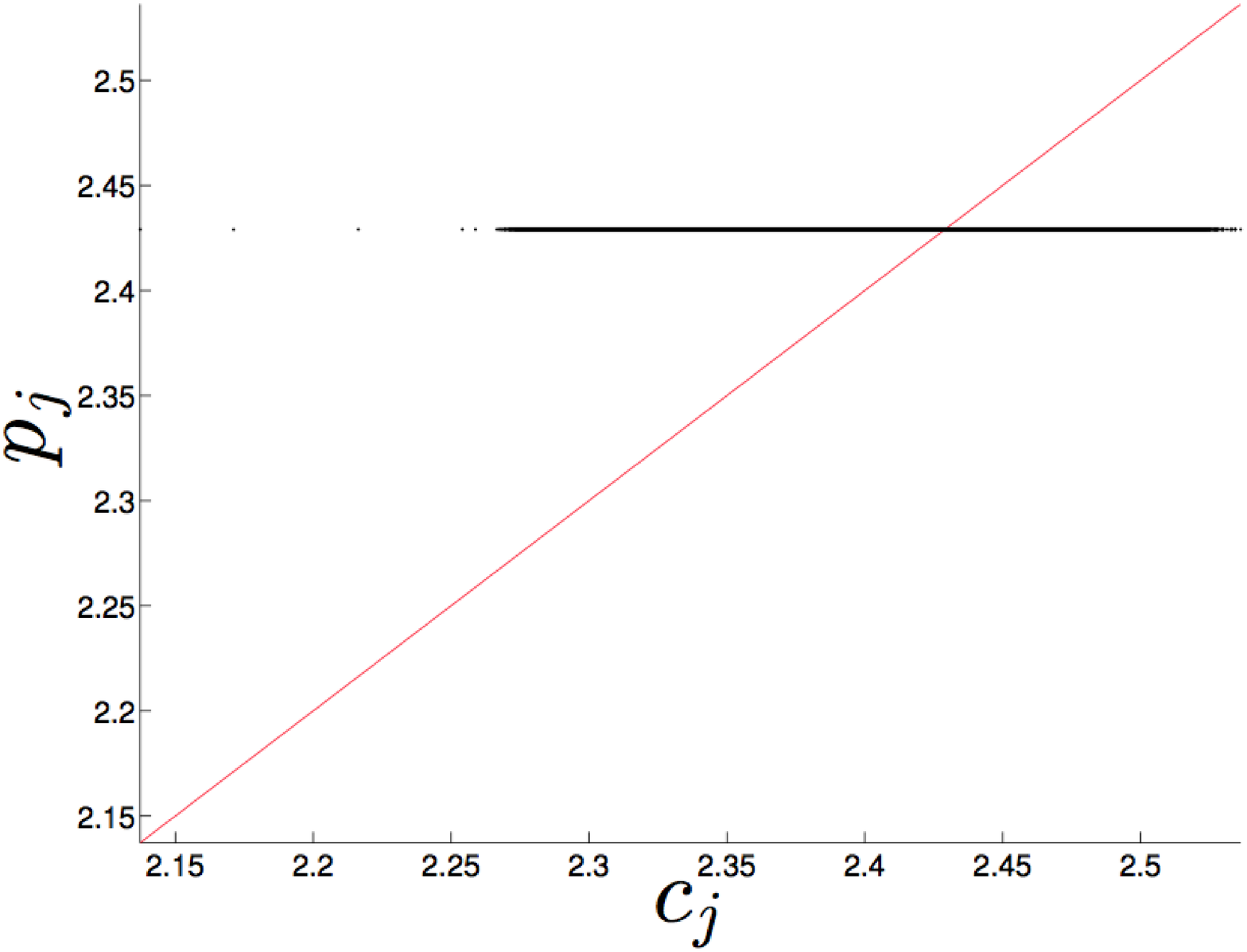} &
    \includegraphics[width=0.6\columnwidth]{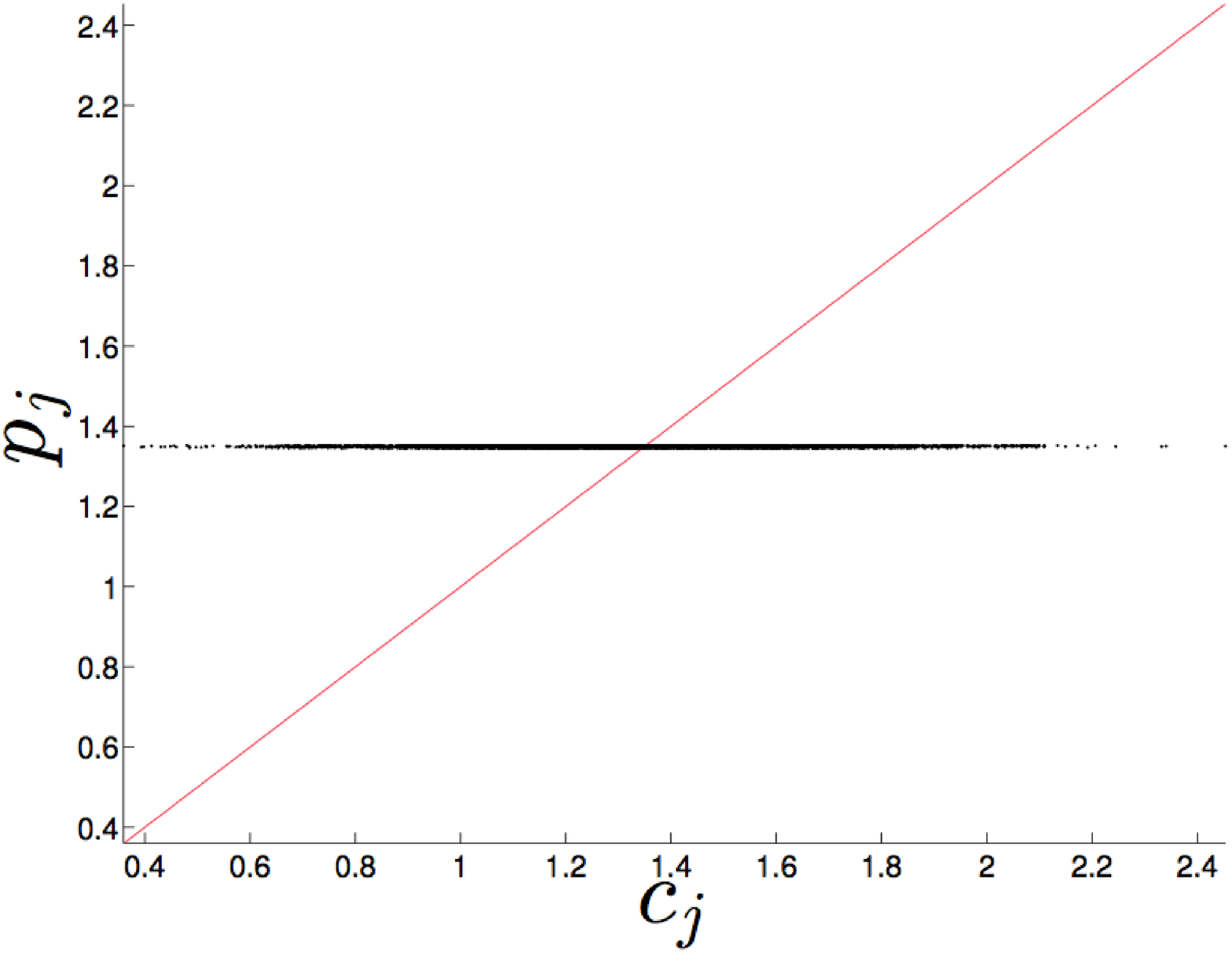} &
    \includegraphics[width=0.6\columnwidth]{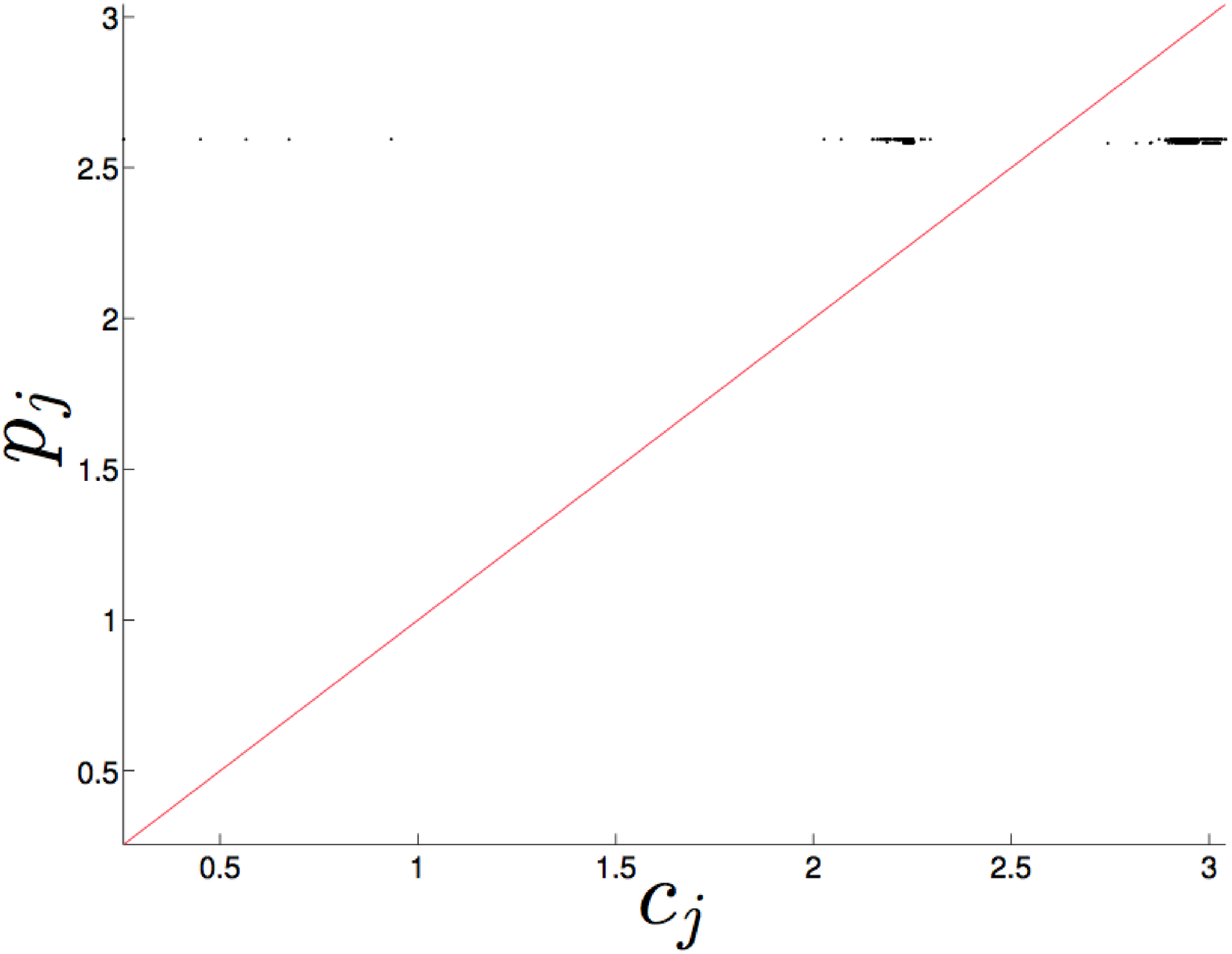} \\
    \rotatebox{90}{\large \arima} &
    \includegraphics[width=0.6\columnwidth]{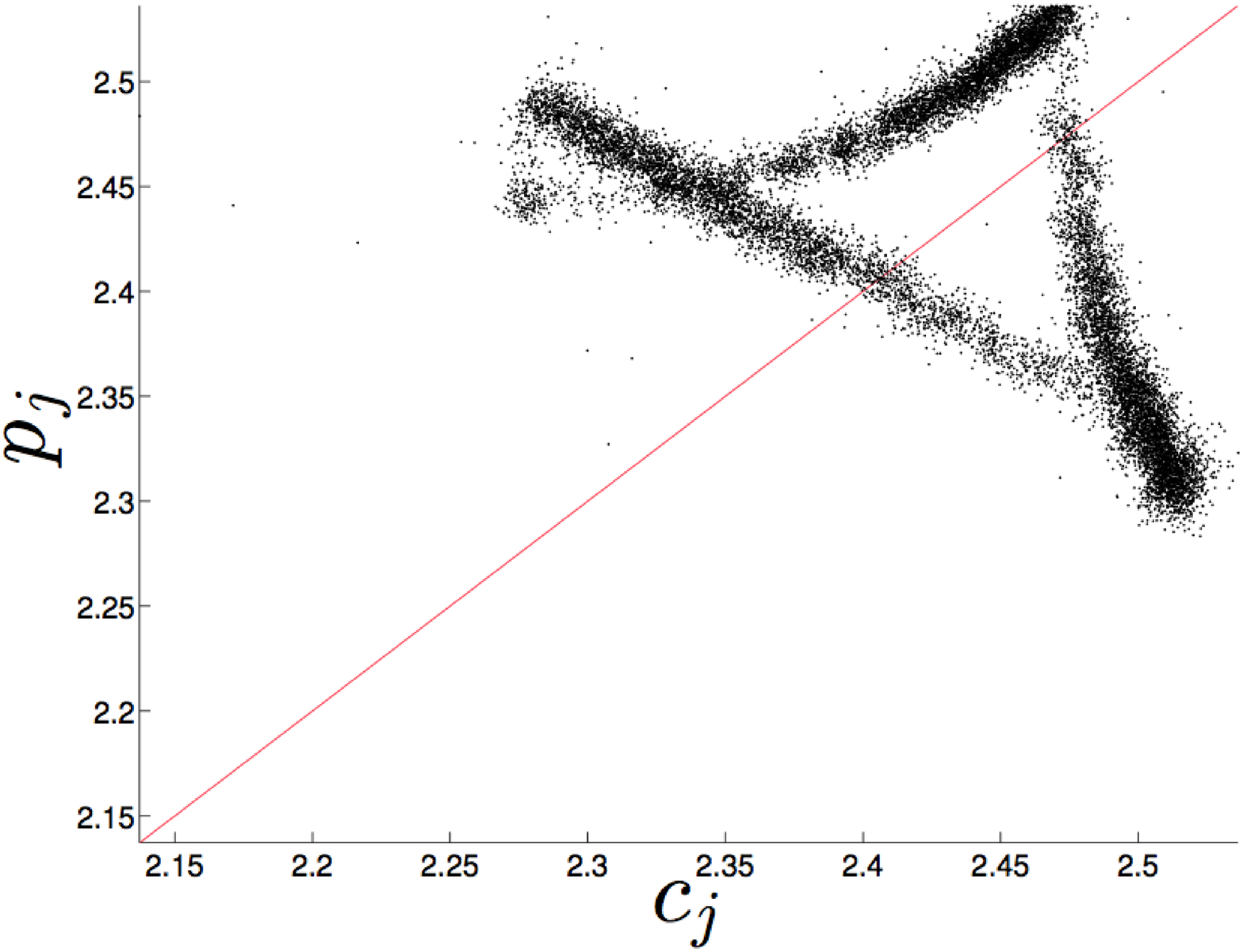} &
    \includegraphics[width=0.6\columnwidth]{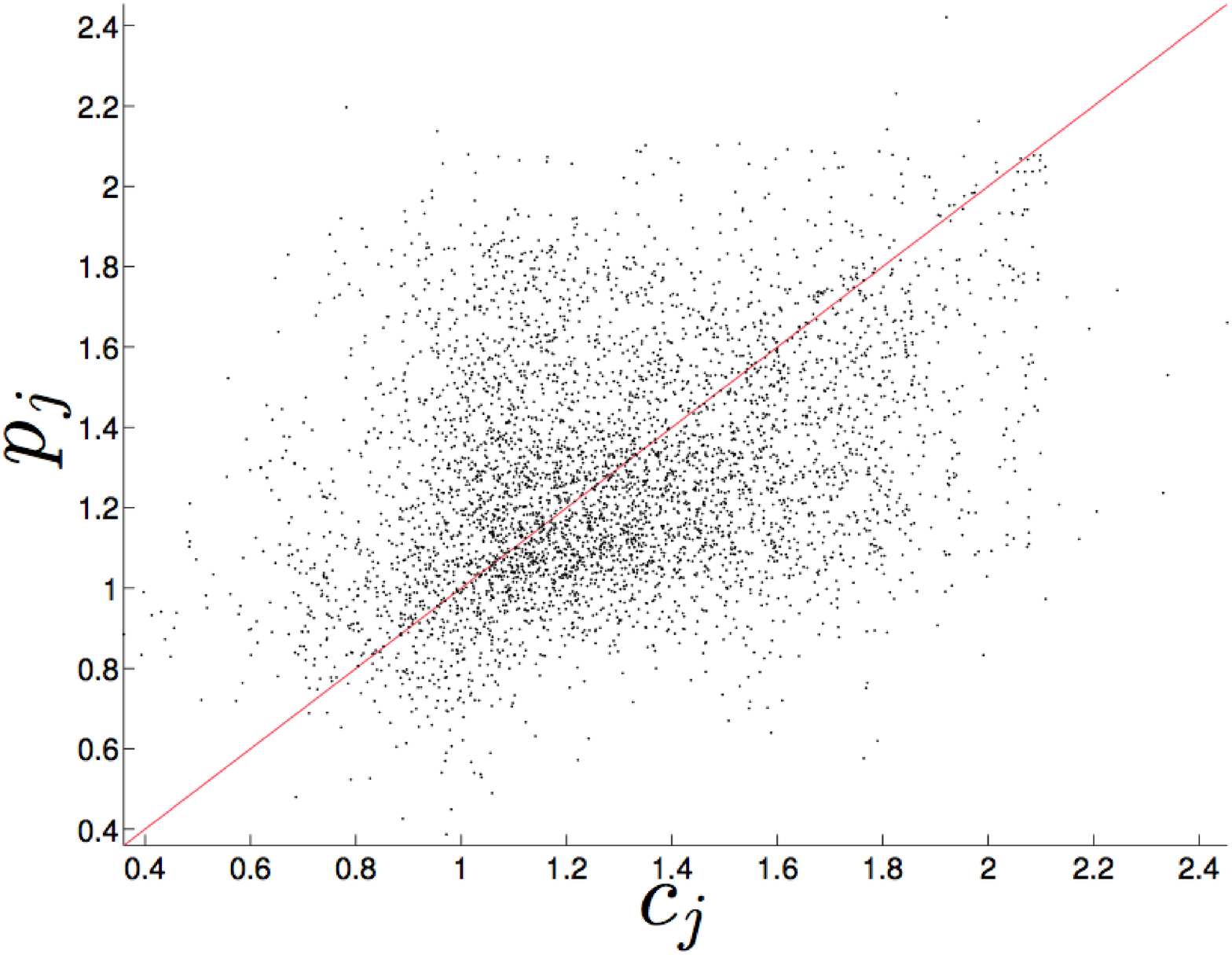} &
    \includegraphics[width=0.6\columnwidth]{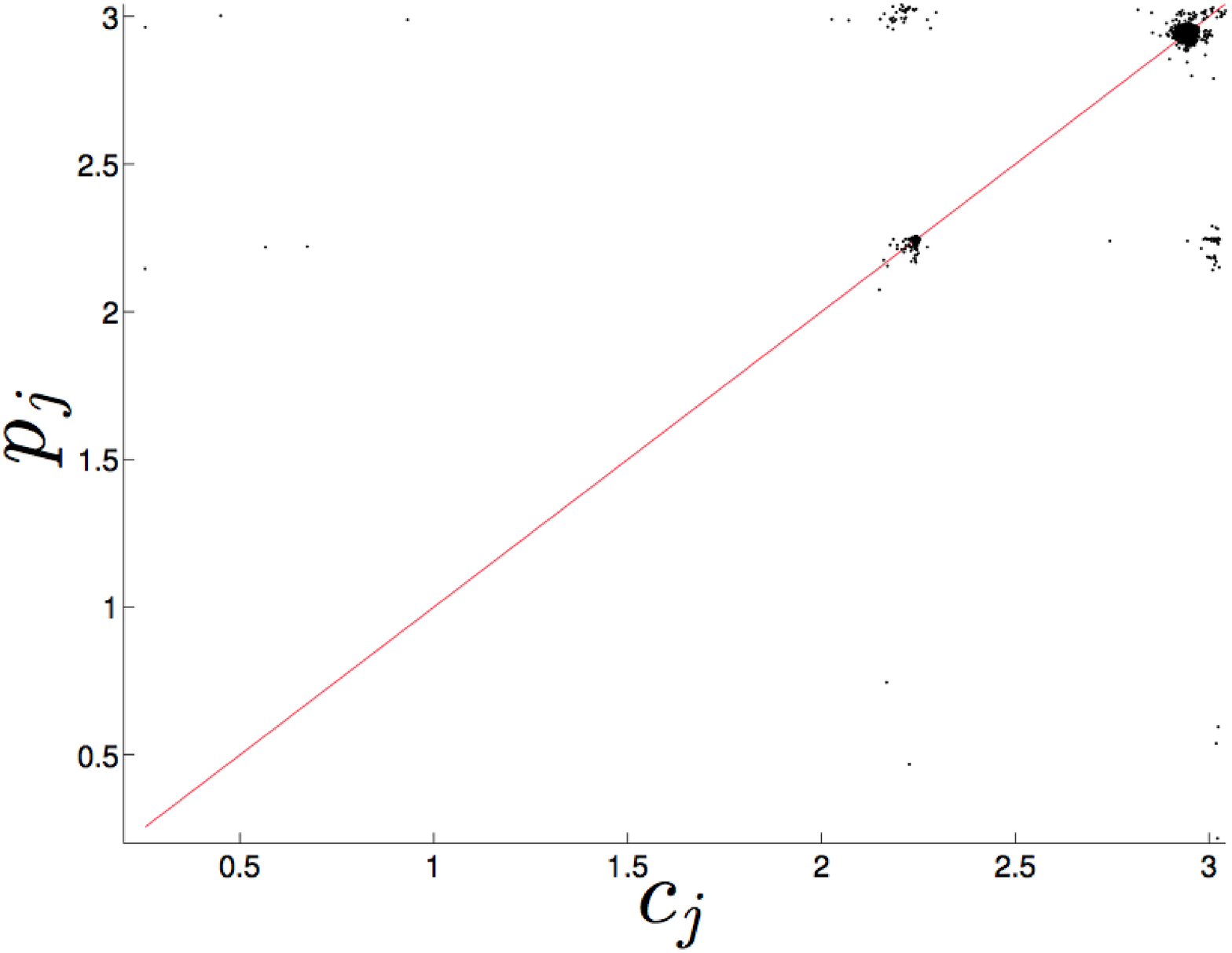} \\
    \rotatebox{90}{\large LMA} &
    \includegraphics[width=0.6\columnwidth]{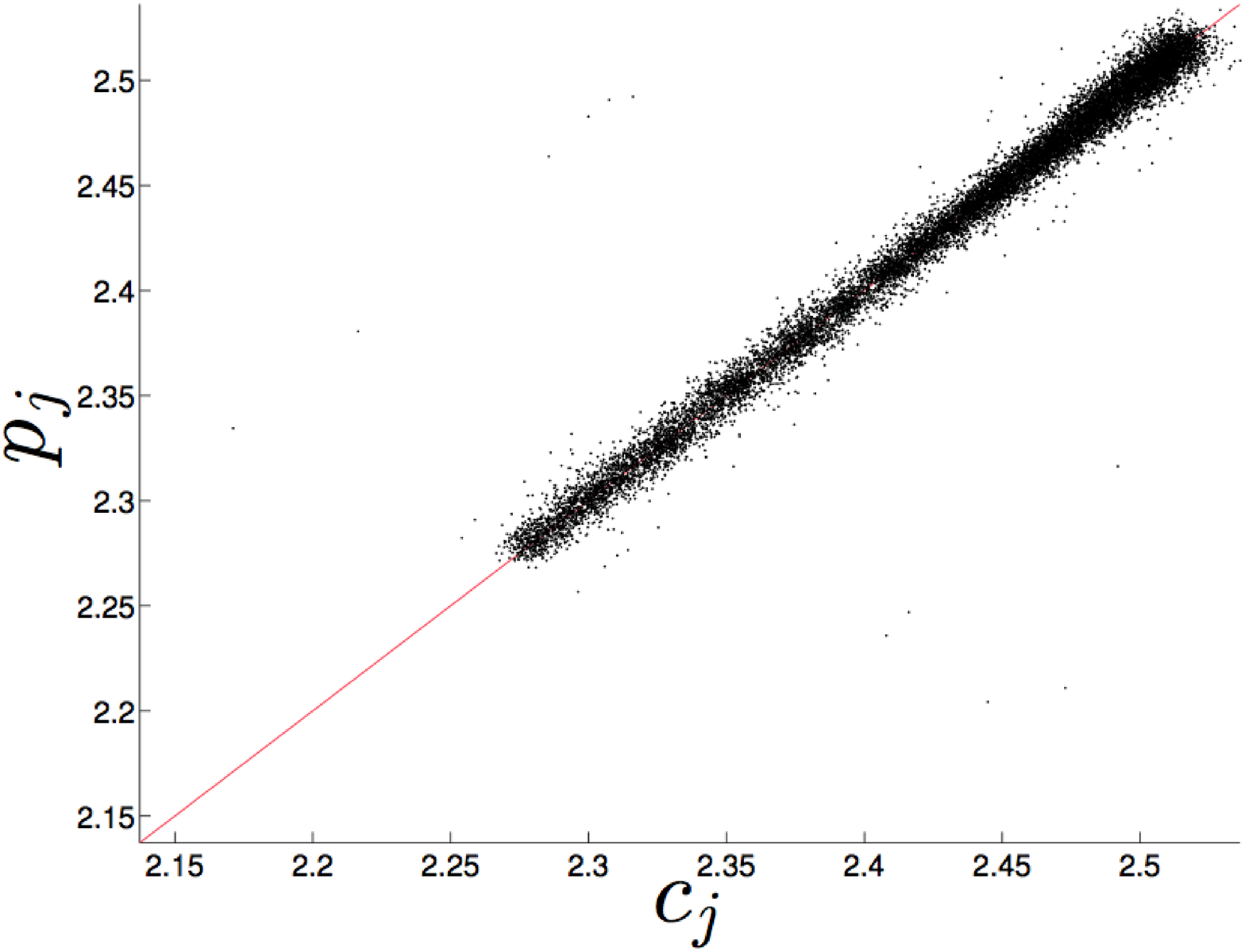} &
    \includegraphics[width=0.6\columnwidth]{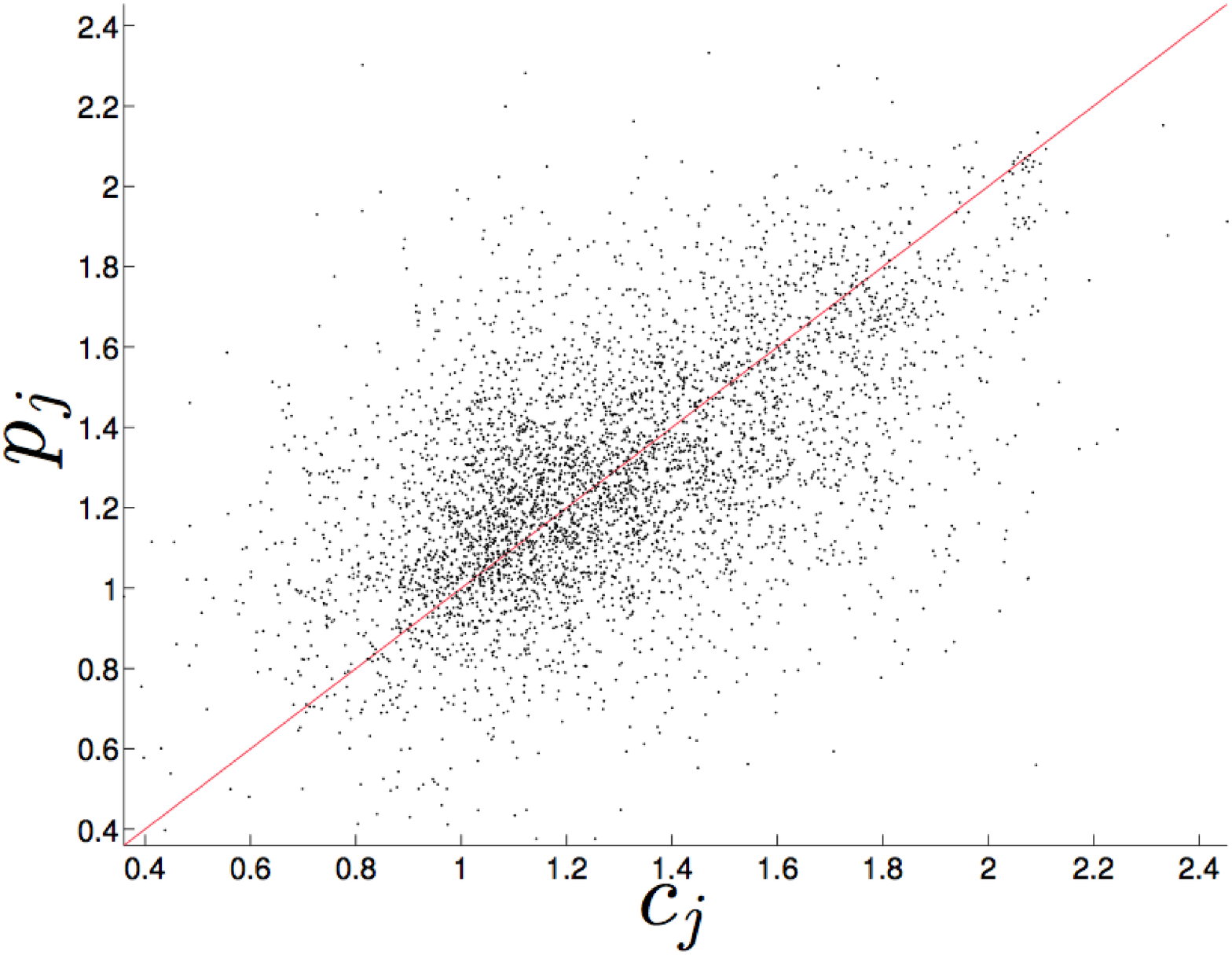} &
    \includegraphics[width=0.6\columnwidth]{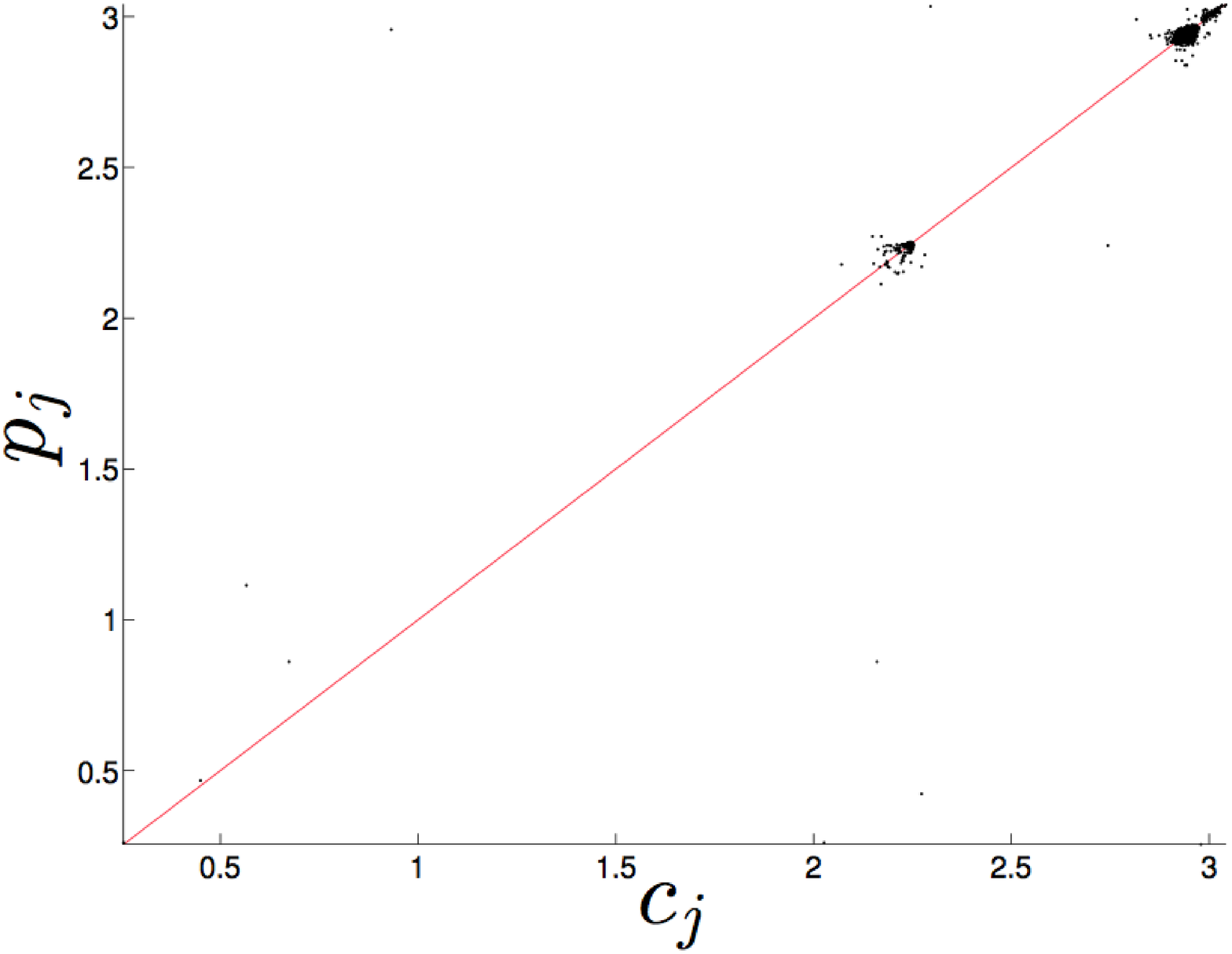}
  \end{tabular}
  \caption{Predicted ($p_j$) versus true values ($c_j$) for forecasts
    of \col, \gcc, and \svdfive made using each of the four 
    strategies studied here.  }
  \label{fig:forecast-example}
\end{figure*}

Figure~\ref{fig:forecast-example} shows example forecasts made using
all four methods for the \col, \gcc, and \svdfive time series.
\newcolumntype{C}{>{\centering\arraybackslash} m{0.6\columnwidth} }
In these images, the vertical axis is the prediction $p_j$ and the
horizontal axis is the true continuation $c_j$.  On such a plot, a
perfect prediction would lie on the diagonal.  LMA, for instance,
generates a very accurate prediction of the \col data, while \arima
does not.  Horizontal lines result when a constant predictor like the
\naive ~method is used on a non-constant signal.  Point clouds reflect
the structure of the distribution of the errors---roughly normally
distributed, for example, in the case of LMA on \gcc.  Note that the
shapes of some of the plots in Figure~\ref{fig:forecast-example}
(e.g., random walk and \arima on \col) are reminiscent of the
projected embedding in Figure~\ref{fig:embedding}(a).  Indeed, for a
random-walk predictor, a $p_j$ vs. $c_j$ plot is technically
equivalent to a two-dimensional embedding with $\tau=1$.  For \arima,
the correspondence is not quite as simple, since the $p_j$ values are
linear combinations of a number of past values of the $c_j$, but the
effect is largely the same\footnote{The temporal \emph{ordering} of
  the points on the \arima $p_j$ vs. $c_j$ plot does not match that of
  a projected embedding of the time series, however.}.

As a numerical measure of prediction accurary, we calculate the mean
absolute scaled error (MASE) between the true and predicted signals:
$$MASE = \sum_{j=n+1}^{k+n+1}\frac{|p_j-c_j|
}{\frac{k}{n-1}\sum^n_{i=2}|x_{i}-x_{i-1}|}$$
This error metric was introduced in \cite{MASE} as a ``generally
applicable measurement of forecast accuracy without the problems seen
in the other measurements."  Like many error metrics, MASE is a
normalized measure: the scaling term in the denominator
%
%
is the average in-sample forecast error for a random-walk prediction
over the initial training signal $\{x_i\}^n_{i=1}$.  That is, MASE$<1$
means that the prediction error in question was, on the average,
smaller than the average error of a random-walk forecast on the
training data.  Analogously, MASE$>1$ means that the corresponding
prediction method did \emph{worse}, on average, than the random-walk
method.  While its comparative nature is somewhat different than
traditional metrics like normalized root mean squared error, MASE has
the significant advantage of allowing one to make a fair comparison
across varying methods, prediction horizons, and signal
scales---attributes that are key to any broad study of predictability.

MASE scores for all 360 experiments are tabulated in the four middle
columns in Table~\ref{tab:error}.
%
%
%
%
%
%
%
%
%
%
 \begin{table*}
\caption{Mean absolute scaled error (MASE) scores and weighted
  permutation entropies for all eight processes studied in this paper.
  LMA $=$ Lorenz method of analogues; RW $=$ random-walk prediction.
  }
  \begin{center}
  \begin{tabular*}{\textwidth}{@{\extracolsep{\fill} } cccccc}
  \hline\hline 
Signal & RW MASE & na\"{i}ve MASE & \arima MASE & LMA MASE & WPE \\
\hline
  \col       & $1.001 \pm 0.002$ & $0.571 \pm 0.002$  & $0.599 \pm 0.211$ & $0.050 \pm 0.002$ & $0.513 \pm 0.003$ \\
  \gcc       & $1.138 \pm 0.011$ & $1.797 \pm 0.010$  & $1.837 \pm 0.016$ & $1.530 \pm 0.021$ & $0.943 \pm 0.001$ \\
  \svdone    & $0.933 \pm 0.095$ & $2.676 \pm 4.328$  & $0.714 \pm 0.075$ & $0.827 \pm 0.076$ & $0.957 \pm 0.016$ \\
  \svdtwo    & $1.125 \pm 0.012$ & $3.054 \pm 0.040$  & $2.163 \pm 0.027$ & $1.279 \pm 0.020$ & $0.846 \pm 0.004$ \\
  \svdthree  & $0.707 \pm 0.009$ & $31.386 \pm 0.282$ & $0.713 \pm 0.010$ & $0.619 \pm 0.021$ & $0.716 \pm 0.006$ \\
  \svdfour   & $1.034 \pm 0.035$ & $2.661 \pm 0.074$  & $0.979 \pm 0.032$ & $0.779 \pm 0.036$ & $0.825 \pm 0.008$ \\
  \svdfive   & $1.001 \pm 0.047$ & $20.870 \pm 0.192$ & $2.370 \pm 0.051$ & $0.718 \pm 0.048$ & $0.678 \pm 0.007$ \\
  \svdsix    & $1.060 \pm 0.055$ & $2.197 \pm 0.083$  & $1.438 \pm 0.061$ & $0.739 \pm 0.068$ & $0.748 \pm 0.011$ \\  
    \hline\hline
  \end{tabular*}
  \end{center}
 \label{tab:error}
  \end{table*}%
In view of the discussion at the end of the previous paragraph, the
fact that the values in the second-from-left column are not
identically equal to 1.00 may be somewhat surprising.  This can happen
due to differences between in-sample and out-of-sample forecasting;
the last 10\% of \svdthree, for instance---the out-of-sample
signal---was more amenable to random-walk prediction, on the average,
than the first 90\%.  See Section~\ref{sec:results} for a deeper
discussion of this effect.

Comparing the values in Table~\ref{tab:error} with the geometry of the
plots in Figure~\ref{fig:forecast-example}, one can see some obvious
correspondences.  The average LMA MASE score for the \col signals was
$0.050$, for instance, while \arima scored much worse ($0.599$).  That
is, LMA performed roughly 20 times better on \col signals than a
random-walk predictor, while \arima only outperformed random walk by a
factor of 1.7.  This is in accordance with the visual appearance of
the corresponding images in Figure~\ref{fig:forecast-example}.  

In other cases, the correspondence between MASE score and the visual
appearance of these kinds of plots is not so clear cut.  The plots of
LMA predictions of \col and \svdfive both lie near the diagonal, for
instance, but the corresponding MASE scores are very different:
$0.050$ for \col and $0.718$ for \svdfive (resp., 20 and 1.4 times
better than random-walk forecasts of the same signals).  This is a
result of the normalization in the MASE score calculation.  Recall
from Section~\ref{sec:simple} that the random-walk method performs
especially poorly on signals that oscillate rapidly.  \col fits this
description, so the random-walk error on this signal---the denominator
of the four MASE scores in the first row of the Table---is
pathologically high, which skews those scores down.  Random-walk
prediction is very effective for the \svdfive signal, on the other
hand, so the denominator is small and the MASE scores are skewed
upwards.

Normalization is, as is often the case, a double-edged sword.  Here,
it facilitates comparisons across signals of different types and
lengths, but its particular pathologies must be taken into account
when analyzing the results, as discussed at more length in
Section~\ref{sec:results}.

\section{Measuring Structural Complexity }\label{sec:meaComplex}

Estimating the entropy of an arbitrary, real-valued time series is a
significant challenge.  Our approach to this problem, which is the topic of
this section, draws upon methods and results from a variety of fields
including time-series analysis, dynamical systems, and stochastic
processes.

For the purposes of this paper, we view the Shannon entropy---in
particular its growth rate with respect to word length (the
\emph{Shannon entropy rate})---as a measure of complexity and
unpredictability in a time series.  Time-series consisting of i.i.d.
random variables, such as white noise, have maximal entropy rates,
whereas highly structured time-series, for example periodic, have very
low (or zero) entropy rates. A time series with a high entropy rate is
almost completely unpredictable; conversely, one with low entropy rate
is often quite predictable. This can be made more rigorous: Pesin's
relation \cite{pesin77} states that in chaotic dynamical systems, the
Kolmogorov-Sinai (KS) entropy is equal to the sum of the positive
Lyapunov exponents, $\lambda_i$.  The Lyapunov exponents directly
quantify the rate at which nearby states of the system diverge with
time: $\left| \Delta x(t) \right| \approx e^{\lambda t} \left| \Delta
x(0) \right|$.  The faster the divergence, the larger the entropy.
The KS entropy is defined as the supremum of the Shannon entropy rates
of all partitions~\cite{petersen1989}. The partition that achieves
this supremum is the \emph{generating partition} discussed in
Section~\ref{sec:related}.

From a different point of view, we can consider the information (as
measured by the Shannon entropy) contained in a single observable, say
the \emph{present}, of the system. This information can be partitioned
into two components: the information shared with past
observations---e.g., the mutual information between the past and
present---and the information in the present that is not contained in
the past (aka ``the conditional entropy of the present given the
past'').  The first part is known as the \emph{redundancy}:
information in the present that is also in the past.  The second part
is the aforementioned Shannon entropy rate.  It seems obvious that the
more redundancy in a signal, the more predictable it should be.  And
the specific \emph{form} of the redundancy should dictate whether a
particular prediction method will work well or poorly on the
corresponding signal.  A linear method cannot capture or make use of
nonlinear redundancy, for instance.  This issue---which has been
explored extensively using clean data and/or systems where the
generating process is known, but not so much with empirical data from
unknown systems---is central to the claims in this paper and the
discussion in the following section.

Previous approaches to measuring temporal complexity via the Shannon
entropy rate \cite{Shannon1951, mantegna1994linguistic} required
categorical data: $x_i \in \mathcal{S}$ for some finite or countably
infinite \emph{alphabet} $\mathcal{S}$.  Data taken from real-world
systems are, however, effectively\footnote{Measurements from
  finite-precision sensors are discrete, but data from modern
  high-resolution sensors are, for the purposes of entropy
  calculations, effectively continuous.}  real-valued.  To analyze
real-valued data using a method that requires categorical values, one
must discretize the data---typically by binning.  Unfortunately, this
is rarely a good solution, as the binning of the values introduces
spurious dynamics~\cite{bollt2001}.  The field of symbolic dynamics
offers discretization methods that do not disturb the intrinsic
behavior.  These methods are, however, fragile in the face of noise;
worse yet, they require knowledge of the underlying system.  This is
inappropriate in our context, where the target of study is
experimental time-series data.

Bandt and Pompe introduced the \emph{permutation entropy} (PE) as a
``natural complexity measure for time series''
\cite{bandt2002per}. The permutation entropy employs a method of
discretizing real-valued time series that follows the intrinsic
behavior of the system under examination.  It has many advantages,
including being robust to observational noise, and its application
does not require any knowledge of the underlying mechanisms.  Rather
than looking at the statistics of sequences of values, as is done when
computing the Shannon entropy, permutation entropy looks at the
statistics of the \emph{orderings} of sequences of values using
ordinal analysis. Ordinal analysis of a time series is the process of
mapping successive time-ordered elements of a time series to their
value-ordered permutation of the same size.  By way of example, if
$(x_1, x_2, x_3) = (9, 1, 7)$ then its \emph{ordinal pattern},
$\phi(x_1, x_2, x_3)$, is $231$ since $x_2 \leq x_3 \leq x_1$.  The
ordinal pattern of the permutation $(x_1, x_2, x_3) = (9, 7, 1)$ is
$321$.

\begin{mydef}[Permutation Entropy]

  Given a time series $\{x_i\}_{i = 1,\dots,N}$. Define $\mathcal{S}_\ell$ as all $\ell!$ permutations $\pi$ of order $\ell$. For each $\pi \in \mathcal{S}_\ell$ we determine the relative frequency of that permutation occurring in $\{x_i\}_{i = 1,\dots,N}$:
  \begin{align*}
    P(\pi) = \frac{\left|\{i|i \leq N-\ell,\phi(x_{i+1},\dots,x_{i+\ell}) = \pi\}\right|}{N-\ell+1}
  \end{align*}
  where $P(\pi)$ quantifies the probability of an ordinal and
  $|\cdot|$ is set cardinality. The \emph{permutation entropy} of
  order $\ell \ge 2$ is defined as
  \begin{align*}
    H(\ell) = - \sum_{\pi \in \mathcal{S}_\ell} P(\pi) \log_2 P(\pi)
  \end{align*}

\end{mydef}

Notice that $0\le H(\ell) \le \log_2(\ell!)$ \cite{bandt2002per}.
With this in mind, it is common in the literature to normalize
permutation entropy as follows: $\frac{H(\ell)}{\log_2(\ell!)}$.  With
this convention, ``low'' PE is close to 0 and ``high'' PE is close to
1. Finally, it should be noted that the permutation entropy has been
shown to be identical to the Kolmolgorov-Sinai entropy for many large
classes of systems \cite{amigo2012permutation}, as long as
observational noise is sufficiently small. As mentioned before, this
is equal to the Shannon entropy rate of a generating partition of the
system. This transitive chain of equalities, from permutation entropy
to Shannon entropy rate via the KS entropy, allows us to approximate
the redundancy---being the dual of the Shannon entropy rate---of a
signal by $1 - \frac{H(\ell)}{\log_2(\ell!)}$.

Here we will be utilizing a variation of the basic permutation entropy
technique, the \emph{weighted permutation entropy} (WPE), which was
introduced in~\cite{fadlallah2013}.  The intent behind the weighting
is to correct for observational noise that is larger than the trends
in the data, but smaller than the larger-scale features.  Consider,
for example, a signal that switches between two fixed points and
contains some additive noise. The PE of such a signal will be
dominated by the noise about the fixed points, driving it to $\approx
1$---which in some sense hides the fact that the signal is actually
quite structured.  In this situation, the terms of the weighted
permutation entropy are dominated by the switching rather than the
stochastic fluctuations.  The amplitude of the switches plays a key
role in predictability, so weighting the scale of ordinal changes is
important for quantifying predictive structure accurately.  To
accomplish this, WPE takes into account the \emph{weight} of a
permutation:
\begin{align*}
  w(x_{i+1}^\ell) = \frac{1}{\ell} \sum_{j = i + 1}^{i+\ell}
                      \left( x_j - \bar{x}_{i+1}^\ell \right)^2
\end{align*}
where $x_{i+1}^\ell$ is a sequence of values $x_{i+1}, \ldots,
x_{i+\ell}$, and $\bar{x}_{i+1}^\ell$ is the arithmetic mean of
those values.

The weighted probability of a permutation is defined as:
\begin{align*}
  P_w(\pi) = \frac{\displaystyle \sum_{i \le N - \ell} w(x_{i+1}^\ell) \cdot \delta(\phi(x_{i+1}^\ell), \pi) }{\displaystyle \sum_{i \le N - \ell} w(x_{i+1}^\ell)}
\end{align*}
where $\delta(x, y)$ is 1 if $x = y$ and 0 otherwise. Effectively,
this weighted probability enhances permutations that are involved in
``large'' features and de-emphasizes permutations that are small in
amplitude relative to the features of the time series. Using the
standard form of an entropy, the weighted permutation entropy is:
\begin{align*}
  H_w(\ell) = - \sum_{\pi \in \mathcal{S}_\ell} P_w(\pi) \log_2 P_w(\pi),
\end{align*}
which can also be normalized by dividing by $\log_2(\ell!)$, making $0
\le \textrm{WPE} \le 1$.  This normalization is used in all the
results that follow.

In practice, calculating permutation entropy and weighted permutation
entropy involves choosing a good value for the word length $\ell$. The
primary consideration in that choice is that the value be large enough
that forbidden ordinals are discovered, yet small enough that
reasonable statistics over the ordinals are gathered.  If an average
of 100 counts per ordinal is considered to be sufficient, for
instance, then $\ell = \operatornamewithlimits{argmax}_{\hat{\ell}} \{
N \gtrapprox 100 \hat{\ell}! \}$.  In the literature, $3 \le \ell \le
6$ is a standard choice---generally without any formal justification.
In theory, the permutation entropy should reach an asymptote with
increasing $\ell$, but that can require an arbitrarily long time
series. In practice, what one should do is calculate the
\emph{persistent} permutation entropy by increasing $\ell$ until the
result converges, but data length issues can intrude before that
convergence is reached.  We used this approach, which we believe
strikes a good balance between accurate ordinal statistics and
finite-data effects, to choose $\ell$ values for the experiments in
the following section.

 \section{Predictability, Complexity, and Permutation Entropy 
 } 
 \label{sec:results}

In this section, we offer an empirical validation of the two findings
introduced in Section \ref{sec:intro}, namely:

\begin{enumerate}

\item The weighted permutation entropy (WPE) of a noisy real-valued
  time series from an unknown system is correlated with prediction
  accuracy---i.e., the predictable structure in an empirical
  time-series data set can be quantified by its WPE.

\item The relationship between WPE and mean absolute scaled error
  (MASE) is a useful empirical heuristic for identifying mismatches
  between prediction models and time-series data---i.e., when there is
  structure in the data that the model is unable to exploit.

\end{enumerate}




The experiments below involve four different prediction methods
applied to time-series data from eight different systems: {\tt
  col\_major}, {\tt 403.gcc}, and the six different segments of the
{\tt dgesdd} signal in Figure~\ref{fig:svd-ts-colored}.  The objective
of these experiments was to explore how prediction accuracy is related
to WPE.
%
%
Working from the first 90\% of each signal, we generated a prediction
of the last 10\% using the random-walk, \naive, \arima, and LMA
prediction methods, as described in Section~\ref{sec:accuracy}, then
calculated the MASE value of those predictions.  We also calculated
the WPE of each time series using a wordlength chosen via the
procedure described at the end of Section~\ref{sec:meaComplex}.  In
order to assess the run-to-run variability of these results, we
repeated all of these calculations on 15 separate trials: i.e., 15
different runs of each program.

\begin{figure}
  \centering
  \includegraphics[width=1.1\columnwidth]{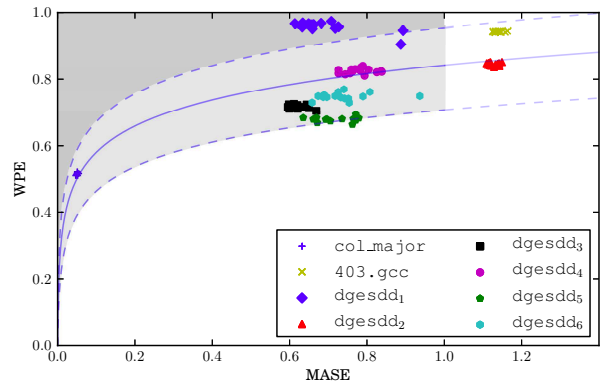}
  \caption{Weighted permutation entropy vs. mean absolute scaled error
    (MASE) of the best prediction of each time series.  The solid
    curve is a least-squares log fit of these points.
%
%
The dashed curves reflect the standard deviation of the model in its
parameter space.  Points that lie below and to the right of the shaded
region indicate that the time series has more predictive structure
than the forecast strategy is able to utilize.}
  \label{fig:wpe_vs_mase_best}
\end{figure}

Figure~\ref{fig:wpe_vs_mase_best} plots the WPE values versus the
corresponding MASE values of the \emph{best} prediction for each of
the 120 time series in this study.
%
%
There is an obvious upward trend, which is consistent with the notion
that there is a pattern in the WPE-MASE relationship.  However, a
simple linear fit is a bad idea here.  First, any signal with zero
entropy should be perfectly predictable (i.e., MASE $\approx 0$), so
any curve fitted to these data should pass through the origin.
Moreover, WPE does not grow without bound, so one would expect the
patterns in the WPE-MASE pairs to reach some sort of asymptote.  For
these reasons, we chose to fit a function of the form $y = a \log(b x
+ 1)$ to these points\footnote{The specific values of the coefficients
  are $a=7.97 \times 10^{-2}$ and $b=1.52 \times 10^3$.}, with $y =$
WPE and $x=$ MASE.  The solid curve in the figure shows this fit; the
dashed curves show the standard deviation of this model in its
parameter space: i.e., $y = a \log(b x + 1)$ with $\pm$ one standard
deviation on each of the two parameters.  Points that fall within this
deviation volume (light grey) correspond to predictions that are
comparable to the best ones found in this study; points that fall
\emph{above} that volume (dark grey) are better still.  We chose to
truncate the shaded region because of a subtle point regarding the
MASE of an ideal predictor, which should not be larger than 1 unless
the training and test signals are different.  This is discussed at
more length below.

The curves and regions in Figure~\ref{fig:wpe_vs_mase_best} are a
graphical representation of the first finding.  This representation
is, we believe, a useful heuristic for determining whether a given
prediction method is well matched to a particular time series.  It is
not, of course, a formal result.  The forecast methods and data sets
used here were chosen to span the space of standard prediction
strategies and the range of dynamical behaviors, but they do not cover
those spaces exhaustively.  Our goal here is an \emph{empirical}
assessment of the relationship between predictability and complexity,
not formal results about a ``best'' predictor for a given time series.
There may be other methods that produce lower MASE values than those
in Figure~\ref{fig:wpe_vs_mase_best}, but the sparseness of the points
above and below the one-$\sigma$ region about the dashed curve in this
plot strongly suggests a pattern of correlation between the underlying
predictability of a time series and its WPE.  The rest of this section
describes these results and claims in more detail---including the
measures taken to assure meaningful comparisons across methods,
trials, and programs---and elaborates on the meaning of the different
curves and limits in the figure.

\begin{figure*}
  \centering
  \includegraphics[width=1.7\columnwidth]{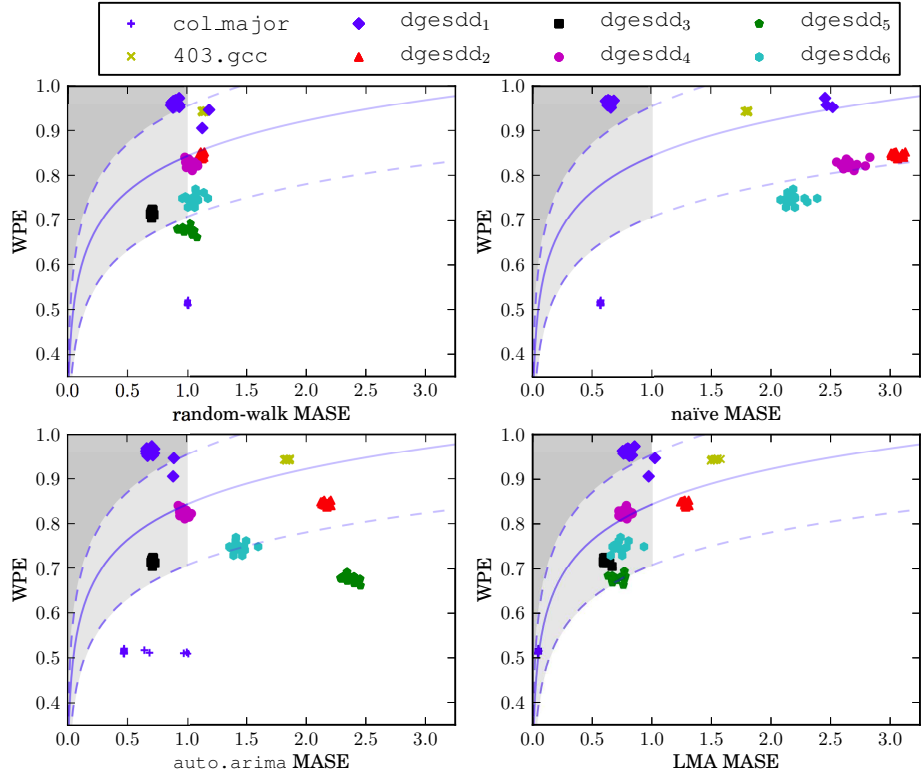}
\caption{WPE vs. MASE for all trials, methods, and systems---with the
  exception of \svdone, \svdthree, and \svdfive are omitted from the
  top-right plot for scale reasons, as described in the text.
%
%
Numerical values, including means and standard deviations of the
errors, can be found in Table~\ref{tab:error}.  The curves and shaded
regions are the same as in the previous figure.  }
    \label{fig:wpe_vs_mase_all}
\end{figure*}

Figure~\ref{fig:wpe_vs_mase_all} shows WPE vs. MASE plots for the full
set of experiments.
There are 15 points in each cluster, one for each trial.  (The points
in Figure~\ref{fig:wpe_vs_mase_best} are the leftmost of the points
for the corresponding trace in any of the four plots in
Figure~\ref{fig:wpe_vs_mase_all}.)  The WPE values do not vary very
much across trials.  For most traces, the variance in MASE scores is
low as well, resulting in small, tight clusters.  In some
cases---\arima predictions of {\tt col\_major}, for instance---the
MASE variance is larger, which spreads out the clusters horizontally.
The mean MASE scores of predictions generated with the nonlinear LMA
method are generally closer to the dashed curve; the \arima method
clusters are more widely spread, the \naive ~clusters even more so.  A
few of the clusters have very high variance; these are discussed later
in this section.

The main thing to note here, however, is not the details of the shapes
of the clusters, but rather their positions in the four plots:
specifically, the fact that many of them are to the right of and/or
below the dashed curve that identifies the boundary of the shaded
region.  \emph{These predictions are not as good as our heuristic
  suggests they could be.}  Focusing in on any single signal makes
this clear: LMA works best for \svdsix, for instance, followed by the
random-walk prediction method, then \arima and \naive.  This provides
some practical leverage: if one calculates an WPE vs. MASE value that
is outside the shaded region, that should suggest that the prediction
method is not well matched to the task at hand: that is, the time
series has more predictive structure than the method is able to use.
%
%
In the case of \arima on \svdsix, for instance, one should try a
different method.  The position of the LMA cluster for \svdsix, on the
other hand, reflects this method's ability to capture and exploit the
structure that is present in this signal.  WPE vs. MASE values like
this, which fall in the shaded region, should suggest to the
practitioner that the prediction method is well-suited to the task.
The following discussion lays out the details that underlie these
claims.
%
%

Though \col is a very simple program, its dynamics are actually quite
complicated, as discussed in Section~\ref{sec:intro}.  Recall from
Figure~\ref{fig:forecast-example} and Table~\ref{tab:error} that the
\naive, \arima, and (especially) random-walk prediction methods do not
perform very well on this signal.  The MASE scores of these
predictions are $0.571 \pm 0.002$, $1.001 \pm 0.002$, and $0.599 \pm
0.211$, respectively, across all 15 trials.  That is, \naive ~and
\arima perform only $\approx 1.7$ times better than the random-walk
method, a primitive strategy that simply uses the current value as the
prediction.  However, the WPE value for the \col trials is $0.513 \pm
0.003$, a signal that is in the center of the complexity spectrum
described in Section~\ref{sec:intro}.

This disparity---WPE values that suggest a high rate of forward
information transfer in the signal, but predictions with comparatively
poor MASE scores---is obvious in the geometry of three of the four
images in Figure~\ref{fig:wpe_vs_mase_all}, where the \col clusters
are far to the right of and/or below the dashed curve.  Again, this
indicates that these methods are not leveraging the available
information in the signal.  The dynamics of \col may be complicated,
but they are not unstructured.  This signal is nonlinear and
deterministic~\cite{mytkowicz09}, and if one uses a prediction
technique that is based a nonlinear model (LMA)---rather than a method
that simply predicts the running mean (\naive) or the previous value
(random walk), or one that uses a linear model (\arima)---the MASE
score is much improved: $0.050 \pm 0.001$.  This prediction is 20
times more accurate than a random-walk forecast, which is more in line
with the level of predictive structure that the low WPE value suggests
is present in the \col signal.  The MASE scores of random-walk
predictions of this signal are all $\approx 1$---as one would
expect---pushing those points well below the shaded region.  Clearly
the stationarity assumption on which this method is based does not
hold for this signal.

The \col example brings out some of the shortcomings of automated
model-building processes.  Note that the {\color{blue}$+$} points are
clustered very tightly in the lower left quadrant of the \naive,
random-walk, and LMA plots in Figure~\ref{fig:wpe_vs_mase_all}, but
spread out horizontally in the \arima plot.  This is because of the
way the \arima process builds ARIMA models \cite{autoARIMA}.  If a
KPSS test of the time series in question indicates that it is
nonstationary, the \arima recipe adds an integration term to the
model.  This test gives mixed results in the case of the \col process,
flagging five of the 15 trials as stationary and ten as nonstationary.
We conjectured that ARIMA models without an integration term perform
more poorly on these five signals, which increases the error and
thereby spreads out the points.  We tested this hypothesis by forcing
the inclusion of an integration term in the five cases where a KPSS
test indicated that such a term was not needed.  This action removed
the spread, pushing all 15 of the \col ~ \arima points in
Figure~\ref{fig:wpe_vs_mase_all} into a tight cluster.

The discussion in the previous paragraph highlights the second finding
of this paper: the ability of our proposed heuristic to flag
inappropriate models.  \arima is an automated, mechanical procedure
for choosing parameters for an ARIMA model of a given data set.  While
the tests and criteria employed by this algorithm
(Section~\ref{sec:arima}) are sophisticated, the results can still be
sub-optimal---if the initial space of models being searched is not
broad enough, for instance, or if one of the preliminary tests gives
an erroneous result.  \arima \emph{always} returns a model, and it can
be very hard to detect when that model is bad.  Our results suggest a
way to do so: if the MASE score of an auto-fitted model like \arima is
out of line with the WPE value of the data, that can be an indication
of inappropriateness in the order selection and parameter estimation
procedure.

The WPE of \svdfive ($0.677 \pm 0.006$) is higher than that of {\tt
  col\_major}.  This indicates that the rate of forward information
transfer of the underlying process is lower, but that time-series data
observed from this system still contain a significant amount of
structure that can, in theory, be used to predict the future course of
the time series.
The MASE scores of the \naive ~and \arima predictions for this system
are $20.870 \pm 0.192$ and $2.370 \pm 0.051$, respectively: that is,
20.87 and 2.37 times worse than a simple random-walk forecast of the
training set portions of the same signals\footnote{The \naive ~MASE
  score is large because of the bimodal nature of the distribution of
  the values of the signal, which makes guessing the mean a
  particularly bad strategy.  The same thing is true of the \svdthree
  signal.}.  As before, the positions of these points on a WPE
vs. MASE plot---significantly below and to the right of the shaded
region---should suggest to a practitioner that
a different method might do better.  Indeed, for \svdfive, the LMA
method produces a MASE score of $ 0.718\pm 0.048 $ and a cluster of
results that largely within the shaded region on the WPE-MASE plot.
This is consistent with our second finding: the LMA method can capture
and reproduce the way in which the \svdfive system processes
information, but the \naive ~and \arima prediction methods cannot.

The WPE of \gcc is higher still: $0.943 \pm 0.001$.  This system
transmits very little information forward in time and provides almost
no structure for prediction methods to work with.  Here, the
random-walk predictor is the best of the methods used here.  This
makes sense; in a fully complex signal, where there is no predictive
structure to utilize, methods that depend on exploiting that
structure---like ARIMA and LMA---cannot get any traction on those
signals.
%
%
Since fitting a hyperplane using least squares should filter out some
of the noise in the signal, the fact that LMA outperforms \arima
($1.530 \pm 0.021$ vs. $1.837 \pm 0.016$) may be somewhat
counterintuitive.  However, the small amount of predictive structure
that is present in this signal is nonlinear (cf.,~\cite{mytkowicz09}),
and LMA is designed to capture and exploit that kind of structure.
Note that all four \gcc clusters in Figure~\ref{fig:wpe_vs_mase_all}
are outside the shaded region; in the case of the random-walk
prediction, for instance, the MASE value is $1.1381 \pm 0.011$.  This
is due to nonstationarity in the signal: in particular, differences
between the training and test signals.  The same effect is at work in
the \svdtwo results, for the same reasons---and visibly so, in the red
segment of Figure~\ref{fig:svd-ts-colored}, where the period and
amplitude of the oscillations are decreasing.

\svdone---the dark blue segment of
Figure~\ref{fig:svd-ts-colored}---behaves very differently than the
other seven systems in this study.  Though its weighted permutation
entropy is very high ($0.957 \pm 0.016$), three of the four prediction
methods do quite well on this signal, yielding mean MASE scores of
0.714 (\arima), 0.827 (LMA), and 0.933 (random walk).  This pushes the
corresponding clusters of points in Figure~\ref{fig:wpe_vs_mase_all}
well above the trend followed by the other seven signals.  The reasons
for this are discussed in the following paragraph.  The MASE scores of
the predictions that were produced by the \naive ~method for this
system, however, are highly inconsistent.  The majority of the blue
diamond-shaped points on the top-right plot in
Figure~\ref{fig:wpe_vs_mase_all} are clustered near a MASE score of
0.6, which is better than the other three methods.  In five of the 15
\svdone trials, however, there were step changes in the signal.  This
is a different nonstationarity than in the case of \col---large jump
discontinuities rather than small shifts in the baseline---and not one
that we were able to handle by simply forcing the ARIMA model to
include a particular term.  The \naive ~method has a very difficult
time with signals like this, particularly if there are multiple step
changes.  This raised the MASE scores of these trials, pushing the
corresponding points to the right\footnote{This includes the cluster
  of three points near MASE $\approx 2.5$, as well as two points that
  are beyond the domain of the graph, at MASE $\approx 11.2-14.8$.},
and in turn raising both the mean and variance of this set of trials.

\begin{figure}[htbp]
  \centering
    \includegraphics[width=\columnwidth]{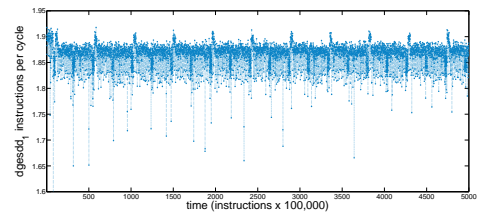}
\caption{A small portion of the \svdone time series}\label{fig:svdone-ts}
\end{figure}

The effects described in the previous paragraph are also exacerbated
by the way MASE is calculated.  Recall that MASE scores are scaled
\emph{relative to a random-walk forecast}.  There are two issues here.
First, random-walk prediction works very badly on signals with
frequent, large, rapid transitions.  Consider a signal that oscillates
from one end of its range to the other at every step.  A signal like
this will have a low WPE, much like {\tt col\_major}.  However, a
random-walk forecast of this signal will be 180 degrees out of phase
with the true continuation.  Since random-walk error appears in the
denominator of the MASE score, this effect can shift points leftwards
on a WPE vs. MASE plot, and that is exactly why the \svdone clusters
in Figure~\ref{fig:wpe_vs_mase_all} are above the dashed curve.  This
time series, part of which is shown in closeup in
Figure~\ref{fig:svdone-ts},
is not quite to the level of the worst-case signal described above,
but it still poses a serious challenge to random-walk prediction.  It
is dominated by a noisy regime (between $\approx$1.86 and
$\approx$1.88 on the vertical scale in Figure~\ref{fig:svdone-ts}),
punctuated by short excursions above 1.9.  In the former regime, which
makes up more than 80\% of the signal, there are frequent dips to 1.82
and occasional larger dips below 1.8.  These single-point dips are the
bane of random-walk forecasting.  In this particular case, roughly
40\% of the forecasted points are off by the width of the associated
dip, which skews the associated MASE scores.  Signals like this are
also problematic for the \naive ~prediction strategy, since the
outliers have significant influence on the mean.  This compounds the
effect of the skew in the scaling factor and exacerbates the spread in
the \svdone MASE values.


The second effect that can skew MASE scores is nonstationarity.  Since
this metric is normalized by the error of a random-walk forecast
\emph{on the training signal}, differences between the test signal and
training signal can create issues.  This is why the MASE values in
Table~\ref{tab:error} are not identically one for every random-walk
forecast of every time series: the last 10\% of these signals is
significantly different from the first 90\%.  The deviation from 1.00
will depend on the process---whether it has multiple regimes, what
those regimes look like, and how it switches between them---as well as
the experimental setup (e.g., sensor precision and data length).  For
the processes studied here, these effects do not cause the MASE values
to exceed 1.15, but pathological situations (e.g., a huge switch in
scale right at the training/test signal boundary, or a signal that
simply grows exponentially) could produce higher values.  This
suggests another potentially useful heuristic: if the MASE of a
random-walk prediction of a time series is significantly different
from 1, this could be an indication that the signal is nonstationary.
We are in the process of exploring this idea and its relationship to
the DVS graphs mentioned in the fourth paragraph of
Section~\ref{sec:related}.

The curves in Figures~\ref{fig:wpe_vs_mase_best}
and~\ref{fig:wpe_vs_mase_all} were determined from finite sets of
methods and data.  We put a lot of thought and effort into making
these sets representative and comprehensive.  The forecast methods
involved range from the simple to the sophisticated; the time-series
data analyzed in this section was sampled from a system whose behavior
spans the dynamical behavior space.  While we are cautiously
optimistic about the generality of our conclusions, more exploration
will be required before we can make any such claim.  Our preliminary
work along those lines shows that data from the H\'enon map
\cite{henon}, the Lorenz system \cite{lorenz}, the SFI A data
set\footnote{This dataset can be retrieved from
  \url{http://www-psych.stanford.edu/~andreas/Time-Series/SantaFe.html}.}
\cite{weigend-book}, and a random-walk process all fall within the
one-$\sigma$ volume of the fit in Figures~\ref{fig:wpe_vs_mase_best}
and~\ref{fig:wpe_vs_mase_all} region, as do various nonlinear
transformations of \svdtwo, \svdfive and \svdsix.

Of course, the geometry of the curves and bounds in these figures do
not necessarily extend to other error metrics.  The positions of the
WPE vs. error points---and the reasons to choose a particular function
to fit to them or impute theoretical bounds on the meaning of the
results---may change if one calculates error using a procedure other
than the one described in Section~\ref{sec:accuracy}.  The MASE error
metric, as discussed above, has its weaknesses.  Even so, we concur
with \cite{MASE} that it is an effective way to compare prediction
error across time-series data of different lengths and different
scales, and we are convinced by the extensive evaluations that are
offered in that paper, as well as the comparisons of MASE to other
metrics, that its strengths far outweigh its weaknesses.

\section{ Conclusions \& Future Work }\label{sec:conc}


Forecast strategies that are designed to capture predictive structure
are ineffective when signal complexity outweighs information
redundancy.  This poses a number of serious challenges in practice.
Without knowing anything about the generating process, it is difficult
to determine how much predictive structure is present in a noisy,
real-world time series.  And even if predictive structure exists, a
given forecast method may not work, simply because it cannot exploit
the structure that is present (e.g., a linear model of a nonlinear
process).  If a forecast model is not producing good results, a
practitioner needs to know why: is the reason that the data contain no
predictive structure---i.e., that no model will work---or is the model
that s/he is using simply not good enough?

In this paper, we have argued that redundancy is a useful proxy for
the inherent predictability of an empirical time series.  To
operationalize that relationship, we use an approximation of the
Kolmogorov-Sinai entropy, estimated using a weighted version of the
permutation entropy of~\cite{bandt2002per}.  This WPE technique---an
ordinal calculation of forward information transfer in a time
series---is ideal for our purposes because it works with real-valued
data and is known to converge to the true entropy value. Using a
variety of forecast models and more than 150 time-series data sets
from experiments and simulations, we have shown that prediction
accuracy is indeed correlated with weighted permutation entropy: the
higher the WPE, in general, the higher the prediction error.  The
relationship is roughly logarithmic, which makes theoretical sense,
given the nature of WPE, predictability, and MASE.

An important practical corollary to this empirical correlation of
predictability and WPE is a practical strategy for assessing
appropriateness of forecast methods.  If the forecast produced by a
particular method is poor but the time series contains a significant
amount of predictive structure, one can reasonably conclude that that
method is inadequate to the task and that one should seek another
method.  The nonlinear LMA method, for instance, performs better in
most cases because it is more general.  (This is particularly apparent
in the \col and \svdfive examples.)
The \naive ~method, which simply predicts the mean, can work very well
on noisy signals because it effects a filtering operation.  The simple
random-walk strategy outperforms LMA, \arima, and the \naive
~method on the \gcc signal, which is extremely complex---i.e.,
extremely low redundancy.

The curves and shaded regions in Figures~\ref{fig:wpe_vs_mase_best}
and~\ref{fig:wpe_vs_mase_all} generalize and operationalize the
discussion in the previous paragraph.  These geometric features are a
preliminary, but potentially useful, heuristic for knowing when a
model is not well-matched to the task at hand: a point that is below
and/or to the right of the shaded regions on a plot like
Figure~\ref{fig:wpe_vs_mase_all} indicates that the time series has
more predictive structure than the forecast model can capture and
exploit---and that one would be well advised to try another method.

These curves were determined empirically using a specific error metric
and a finite set of forecast methods and time-series traces.  If one
uses a different error metric, the geometry of the heuristic will be
different---and may not even make sense, if one uses a metric that
does not support comparison across different time series.  And while
the methods and traces used in this study were chosen to be
representative of the practice, they are of course not completely
comprehensive.  It is certainly possible, for instance, that the
nonlinear dynamics of computer performance is subtly different from
the nonlinear dynamics of other systems.  Our preliminary results on
other systems (H\'enon, Lorenz, a random-walk process, SFI ``A'',
nonlinear transformations of the computer performance data) lead us to believe
that our results will generalize beyond the examples described in this
paper.  We are in the process of following up on that exploration with
a broader study of data, forecast methods, and error metrics.

%


Nonstationarity is a serious challenge in any time-series modeling
problem.  Any regime shift that causes a change in the predictive
structure between the training signal and the test signal, for
instance, may skew the MASE score and thereby affecting the utility of
our heuristic.  Conversely, though, a MASE score of a random-walk
prediction that is significantly different from 1.0---as mentioned at
end of the previous section---could potentially be a good indicator of
nonstationarity.

Detecting regime shifts---and adapting prediction models
accordingly---is an important area of future work.  Indeed, one of the
first applications of permutation entropy was to recognize the regime
shift in brainwave data that occurs when someone has a
seizure~\cite{cao2004det}.  Recall that the signal in
Figure~\ref{fig:svd-ts-colored} was especially useful for the study in
this paper because it contained a number of different regimes.  We
segmented this signal visually, but one could imagine using some
combination of WPE and MASE to do so instead (e.g., in a sliding
window across the time series).  Automating regime-shift detection
would be an important step towards a fully adaptive modeling strategy,
where old models are discarded and new ones are rebuilt whenever the
time series enters a new regime.  Our WPE vs. MASE results could be
particularly powerful in this scenario, as their values could not only
help with regime-shift detection, but also suggest what kind of model
might work well in each new regime.  Of particular interest would be
the class of so-called \emph{hybrid systems}~\cite{hybrid}, which
exhibit discrete transitions between different continuous
regimes---e.g., a lathe that has an intermittent instability or
traffic at an internet router, whose characteristic normal traffic
patterns shift radically during an attack.  Effective modeling and
prediction of these kinds of systems is quite difficult; doing so
adaptively and automatically---in the manner that is alluded to at the
end of the previous paragraph---would be an interesting challenge.

\section*{Acknowledgment}
This work was partially supported by NSF grant \#CMMI-1245947 and ARO
grant \#W911NF-12-1-0288.  We would also like to acknowledge the Santa
Fe Institute whose interdisciplinary halls created a place for this
work to be born, and the anonymous reviewers whose comments
strengthened the presentation.

\bibliographystyle{apsrev4-1}
\bibliography{bibliofile}

\begin{thebibliography}{48}%
\makeatletter
\providecommand \@ifxundefined [1]{%
 \@ifx{#1\undefined}
}%
\providecommand \@ifnum [1]{%
 \ifnum #1\expandafter \@firstoftwo
 \else \expandafter \@secondoftwo
 \fi
}%
\providecommand \@ifx [1]{%
 \ifx #1\expandafter \@firstoftwo
 \else \expandafter \@secondoftwo
 \fi
}%
\providecommand \natexlab [1]{#1}%
\providecommand \enquote  [1]{``#1''}%
\providecommand \bibnamefont  [1]{#1}%
\providecommand \bibfnamefont [1]{#1}%
\providecommand \citenamefont [1]{#1}%
\providecommand \href@noop [0]{\@secondoftwo}%
\providecommand \href [0]{\begingroup \@sanitize@url \@href}%
\providecommand \@href[1]{\@@startlink{#1}\@@href}%
\providecommand \@@href[1]{\endgroup#1\@@endlink}%
\providecommand \@sanitize@url [0]{\catcode `\\12\catcode `\$12\catcode
  `\&12\catcode `\#12\catcode `\^12\catcode `\_12\catcode `\%12\relax}%
\providecommand \@@startlink[1]{}%
\providecommand \@@endlink[0]{}%
\providecommand \url  [0]{\begingroup\@sanitize@url \@url }%
\providecommand \@url [1]{\endgroup\@href {#1}{\urlprefix }}%
\providecommand \urlprefix  [0]{URL }%
\providecommand \Eprint [0]{\href }%
\providecommand \doibase [0]{http://dx.doi.org/}%
\providecommand \selectlanguage [0]{\@gobble}%
\providecommand \bibinfo  [0]{\@secondoftwo}%
\providecommand \bibfield  [0]{\@secondoftwo}%
\providecommand \translation [1]{[#1]}%
\providecommand \BibitemOpen [0]{}%
\providecommand \bibitemStop [0]{}%
\providecommand \bibitemNoStop [0]{.\EOS\space}%
\providecommand \EOS [0]{\spacefactor3000\relax}%
\providecommand \BibitemShut  [1]{\csname bibitem#1\endcsname}%
\let\auto@bib@innerbib\@empty
\bibitem [{\citenamefont {Crutchfield}\ and\ \citenamefont
  {Feldman}(2003)}]{crutchfield2003}%
  \BibitemOpen
  \bibfield  {author} {\bibinfo {author} {\bibfnamefont {J.~P.}\ \bibnamefont
  {Crutchfield}}\ and\ \bibinfo {author} {\bibfnamefont {D.~P.}\ \bibnamefont
  {Feldman}},\ }\href@noop {} {\bibfield  {journal} {\bibinfo  {journal}
  {Chaos}\ }\textbf {\bibinfo {volume} {13}},\ \bibinfo {pages} {25} (\bibinfo
  {year} {2003})}\BibitemShut {NoStop}%
\bibitem [{\citenamefont {Lind}\ and\ \citenamefont {Marcus}(1995)}]{lind95}%
  \BibitemOpen
  \bibfield  {author} {\bibinfo {author} {\bibfnamefont {D.}~\bibnamefont
  {Lind}}\ and\ \bibinfo {author} {\bibfnamefont {B.}~\bibnamefont {Marcus}},\
  }\href@noop {} {\emph {\bibinfo {title} {An Introduction to Symbolic Dynamics
  and Coding}}}\ (\bibinfo  {publisher} {Cambridge University Press},\ \bibinfo
  {year} {1995})\BibitemShut {NoStop}%
\bibitem [{\citenamefont {Shalizi}\ and\ \citenamefont
  {Crutchfield}(2001)}]{Shalizi2008}%
  \BibitemOpen
  \bibfield  {author} {\bibinfo {author} {\bibfnamefont {C.~R.}\ \bibnamefont
  {Shalizi}}\ and\ \bibinfo {author} {\bibfnamefont {J.~P.}\ \bibnamefont
  {Crutchfield}},\ }\href@noop {} {\bibfield  {journal} {\bibinfo  {journal}
  {J. Stat. Phys.}\ }\textbf {\bibinfo {volume} {104}},\ \bibinfo {pages} {817}
  (\bibinfo {year} {2001})}\BibitemShut {NoStop}%
\bibitem [{\citenamefont {Bandt}\ and\ \citenamefont
  {Pompe}(2002)}]{bandt2002per}%
  \BibitemOpen
  \bibfield  {author} {\bibinfo {author} {\bibfnamefont {C.}~\bibnamefont
  {Bandt}}\ and\ \bibinfo {author} {\bibfnamefont {B.}~\bibnamefont {Pompe}},\
  }\href@noop {} {\bibfield  {journal} {\bibinfo  {journal} {Phys. Rev. Lett.}\
  }\textbf {\bibinfo {volume} {88}},\ \bibinfo {pages} {174102} (\bibinfo
  {year} {2002})}\BibitemShut {NoStop}%
\bibitem [{\citenamefont {Bollt}\ \emph {et~al.}(2001)\citenamefont {Bollt},
  \citenamefont {Stanford}, \citenamefont {Lai},\ and\ \citenamefont
  {{\.Z}yczkowski}}]{bollt2001}%
  \BibitemOpen
  \bibfield  {author} {\bibinfo {author} {\bibfnamefont {E.~M.}\ \bibnamefont
  {Bollt}}, \bibinfo {author} {\bibfnamefont {T.}~\bibnamefont {Stanford}},
  \bibinfo {author} {\bibfnamefont {Y.-C.}\ \bibnamefont {Lai}}, \ and\
  \bibinfo {author} {\bibfnamefont {K.}~\bibnamefont {{\.Z}yczkowski}},\
  }\href@noop {} {\bibfield  {journal} {\bibinfo  {journal} {Phys. D}\ }\textbf
  {\bibinfo {volume} {154}},\ \bibinfo {pages} {259} (\bibinfo {year}
  {2001})}\BibitemShut {NoStop}%
\bibitem [{\citenamefont {Yule}(1927)}]{Yule27}%
  \BibitemOpen
  \bibfield  {author} {\bibinfo {author} {\bibfnamefont {U.~G.}\ \bibnamefont
  {Yule}},\ }\href@noop {} {\bibfield  {journal} {\bibinfo  {journal} {Phil.
  Trans. R. Soc. London A}\ }\textbf {\bibinfo {volume} {226}},\ \bibinfo
  {pages} {267} (\bibinfo {year} {1927})}\BibitemShut {NoStop}%
\bibitem [{\citenamefont {Smith}(1992)}]{Smith199250}%
  \BibitemOpen
  \bibfield  {author} {\bibinfo {author} {\bibfnamefont {L.~A.}\ \bibnamefont
  {Smith}},\ }\href@noop {} {\bibfield  {journal} {\bibinfo  {journal} {Phys.
  D}\ }\textbf {\bibinfo {volume} {58}},\ \bibinfo {pages} {50 } (\bibinfo
  {year} {1992})}\BibitemShut {NoStop}%
\bibitem [{\citenamefont {Gershenfeld}\ and\ \citenamefont
  {Weigend}(1993)}]{weigend93}%
  \BibitemOpen
  \bibfield  {author} {\bibinfo {author} {\bibfnamefont {N.}~\bibnamefont
  {Gershenfeld}}\ and\ \bibinfo {author} {\bibfnamefont {A.}~\bibnamefont
  {Weigend}},\ }in\ \href@noop {} {\emph {\bibinfo {booktitle} {Time Series
  Prediction: Forecasting the Future and Understanding the Past}}},\ \bibinfo
  {editor} {edited by\ \bibinfo {editor} {\bibfnamefont {A.}~\bibnamefont
  {Weigend}}\ and\ \bibinfo {editor} {\bibfnamefont {N.}~\bibnamefont
  {Gershenfeld}}}\ (\bibinfo  {publisher} {Addison-Wesley},\ \bibinfo {year}
  {1993})\BibitemShut {NoStop}%
\bibitem [{\citenamefont {Casdagli}(1992)}]{Casdagli92dvsplots}%
  \BibitemOpen
  \bibfield  {author} {\bibinfo {author} {\bibfnamefont {M.}~\bibnamefont
  {Casdagli}},\ }\href@noop {} {\bibfield  {journal} {\bibinfo  {journal} {J.
  R. Stat. Soc. B}\ }\textbf {\bibinfo {volume} {54}},\ \bibinfo {pages} {303}
  (\bibinfo {year} {1992})}\BibitemShut {NoStop}%
\bibitem [{\citenamefont {Shannon}(1951)}]{Shannon1951}%
  \BibitemOpen
  \bibfield  {author} {\bibinfo {author} {\bibfnamefont {C.~E.}\ \bibnamefont
  {Shannon}},\ }\href@noop {} {\bibfield  {journal} {\bibinfo  {journal} {Bell
  Syst. Tech. J.}\ }\textbf {\bibinfo {volume} {30}},\ \bibinfo {pages} {50}
  (\bibinfo {year} {1951})}\BibitemShut {NoStop}%
\bibitem [{\citenamefont {Mantegna}\ \emph {et~al.}(1994)\citenamefont
  {Mantegna}, \citenamefont {Buldyrev}, \citenamefont {Goldberger},
  \citenamefont {Havlin}, \citenamefont {Peng}, \citenamefont {Simons},\ and\
  \citenamefont {Stanley}}]{mantegna1994linguistic}%
  \BibitemOpen
  \bibfield  {author} {\bibinfo {author} {\bibfnamefont {R.~N.}\ \bibnamefont
  {Mantegna}}, \bibinfo {author} {\bibfnamefont {S.~V.}\ \bibnamefont
  {Buldyrev}}, \bibinfo {author} {\bibfnamefont {A.~L.}\ \bibnamefont
  {Goldberger}}, \bibinfo {author} {\bibfnamefont {S.}~\bibnamefont {Havlin}},
  \bibinfo {author} {\bibfnamefont {C.~K.}\ \bibnamefont {Peng}}, \bibinfo
  {author} {\bibfnamefont {M.}~\bibnamefont {Simons}}, \ and\ \bibinfo {author}
  {\bibfnamefont {H.~E.}\ \bibnamefont {Stanley}},\ }\href@noop {} {\bibfield
  {journal} {\bibinfo  {journal} {Phys. Rev. Lett.}\ }\textbf {\bibinfo
  {volume} {73}},\ \bibinfo {pages} {3169} (\bibinfo {year}
  {1994})}\BibitemShut {NoStop}%
\bibitem [{\citenamefont {Eisele}(1999)}]{eisele1999}%
  \BibitemOpen
  \bibfield  {author} {\bibinfo {author} {\bibfnamefont {M.}~\bibnamefont
  {Eisele}},\ }\href@noop {} {\bibfield  {journal} {\bibinfo  {journal} {J.
  Phys. A}\ }\textbf {\bibinfo {volume} {32}},\ \bibinfo {pages} {1533}
  (\bibinfo {year} {1999})}\BibitemShut {NoStop}%
\bibitem [{\citenamefont {Haven}\ \emph {et~al.}(2005)\citenamefont {Haven},
  \citenamefont {Majda},\ and\ \citenamefont {Abramov}}]{haven2005}%
  \BibitemOpen
  \bibfield  {author} {\bibinfo {author} {\bibfnamefont {K.}~\bibnamefont
  {Haven}}, \bibinfo {author} {\bibfnamefont {A.}~\bibnamefont {Majda}}, \ and\
  \bibinfo {author} {\bibfnamefont {R.}~\bibnamefont {Abramov}},\ }\href@noop
  {} {\bibfield  {journal} {\bibinfo  {journal} {J. Comput. Phys.}\ }\textbf
  {\bibinfo {volume} {206}},\ \bibinfo {pages} {334} (\bibinfo {year}
  {2005})}\BibitemShut {NoStop}%
\bibitem [{\citenamefont {Boffetta}\ \emph {et~al.}(2002)\citenamefont
  {Boffetta}, \citenamefont {Cencini}, \citenamefont {Falcioni},\ and\
  \citenamefont {Vulpiani}}]{boffetta02}%
  \BibitemOpen
  \bibfield  {author} {\bibinfo {author} {\bibfnamefont {G.}~\bibnamefont
  {Boffetta}}, \bibinfo {author} {\bibfnamefont {M.}~\bibnamefont {Cencini}},
  \bibinfo {author} {\bibfnamefont {M.}~\bibnamefont {Falcioni}}, \ and\
  \bibinfo {author} {\bibfnamefont {A.}~\bibnamefont {Vulpiani}},\ }\href@noop
  {} {\bibfield  {journal} {\bibinfo  {journal} {Phys. Rep.}\ }\textbf
  {\bibinfo {volume} {356}},\ \bibinfo {pages} {367} (\bibinfo {year}
  {2002})}\BibitemShut {NoStop}%
\bibitem [{\citenamefont {Myktowicz}\ \emph {et~al.}(2009)\citenamefont
  {Myktowicz}, \citenamefont {Diwan},\ and\ \citenamefont
  {Bradley}}]{mytkowicz09}%
  \BibitemOpen
  \bibfield  {author} {\bibinfo {author} {\bibfnamefont {T.}~\bibnamefont
  {Myktowicz}}, \bibinfo {author} {\bibfnamefont {A.}~\bibnamefont {Diwan}}, \
  and\ \bibinfo {author} {\bibfnamefont {E.}~\bibnamefont {Bradley}},\
  }\href@noop {} {\bibfield  {journal} {\bibinfo  {journal} {Chaos}\ }\textbf
  {\bibinfo {volume} {19}},\ \bibinfo {eid} {033124} (\bibinfo {year}
  {2009})}\BibitemShut {NoStop}%
\bibitem [{\citenamefont {Hopcroft}\ \emph {et~al.}(2007)\citenamefont
  {Hopcroft}, \citenamefont {Motwani},\ and\ \citenamefont
  {Ullman}}]{hopcroft2007}%
  \BibitemOpen
  \bibfield  {author} {\bibinfo {author} {\bibfnamefont {J.}~\bibnamefont
  {Hopcroft}}, \bibinfo {author} {\bibfnamefont {R.}~\bibnamefont {Motwani}}, \
  and\ \bibinfo {author} {\bibfnamefont {J.}~\bibnamefont {Ullman}},\
  }\href@noop {} {\emph {\bibinfo {title} {Introduction to automata theory,
  languages, and computation}}}\ (\bibinfo  {publisher} {Pearson/Addison
  Wesley},\ \bibinfo {year} {2007})\BibitemShut {NoStop}%
\bibitem [{\citenamefont {Garland}\ and\ \citenamefont
  {Bradley}(2011)}]{josh-ida2011}%
  \BibitemOpen
  \bibfield  {author} {\bibinfo {author} {\bibfnamefont {J.}~\bibnamefont
  {Garland}}\ and\ \bibinfo {author} {\bibfnamefont {E.}~\bibnamefont
  {Bradley}},\ }in\ \href@noop {} {\emph {\bibinfo {booktitle} {Advances in
  Intelligent Data Analysis X}}},\ \bibinfo {series} {Lecture Notes in Computer
  Science}, Vol.\ \bibinfo {volume} {7014},\ \bibinfo {editor} {edited by\
  \bibinfo {editor} {\bibfnamefont {J.}~\bibnamefont {Gama}}, \bibinfo {editor}
  {\bibfnamefont {E.}~\bibnamefont {Bradley}}, \ and\ \bibinfo {editor}
  {\bibfnamefont {J.}~\bibnamefont {Hollm{\'e}n}}}\ (\bibinfo  {publisher}
  {Springer Berlin/Heidelberg},\ \bibinfo {year} {2011})\ pp.\ \bibinfo {pages}
  {173--184}\BibitemShut {NoStop}%
\bibitem [{\citenamefont {Garland}\ and\ \citenamefont
  {Bradley}(2013)}]{josh-ida2013}%
  \BibitemOpen
  \bibfield  {author} {\bibinfo {author} {\bibfnamefont {J.}~\bibnamefont
  {Garland}}\ and\ \bibinfo {author} {\bibfnamefont {E.}~\bibnamefont
  {Bradley}},\ }in\ \href@noop {} {\emph {\bibinfo {booktitle} {Advances in
  Intelligent Data Analysis XII}}},\ \bibinfo {series} {Lecture Notes in
  Computer Science}, Vol.\ \bibinfo {volume} {8207},\ \bibinfo {editor} {edited
  by\ \bibinfo {editor} {\bibfnamefont {A.}~\bibnamefont {Tucker}}, \bibinfo
  {editor} {\bibfnamefont {F.}~\bibnamefont {H{\"o}ppner}}, \bibinfo {editor}
  {\bibfnamefont {A.}~\bibnamefont {Siebes}}, \ and\ \bibinfo {editor}
  {\bibfnamefont {S.}~\bibnamefont {Swift}}}\ (\bibinfo  {publisher} {Springer
  Berlin/Heidelberg},\ \bibinfo {year} {2013})\ pp.\ \bibinfo {pages}
  {210--222}\BibitemShut {NoStop}%
\bibitem [{\citenamefont {Alexander}\ \emph {et~al.}(2010)\citenamefont
  {Alexander}, \citenamefont {Mytkowicz}, \citenamefont {Diwan},\ and\
  \citenamefont {Bradley}}]{zach-IDA10}%
  \BibitemOpen
  \bibfield  {author} {\bibinfo {author} {\bibfnamefont {Z.}~\bibnamefont
  {Alexander}}, \bibinfo {author} {\bibfnamefont {T.}~\bibnamefont
  {Mytkowicz}}, \bibinfo {author} {\bibfnamefont {A.}~\bibnamefont {Diwan}}, \
  and\ \bibinfo {author} {\bibfnamefont {E.}~\bibnamefont {Bradley}},\ }in\
  \href@noop {} {\emph {\bibinfo {booktitle} {Advances in Intelligent Data
  Analysis IX}}},\ \bibinfo {series} {Lecture Notes in Computer Science}, Vol.\
  \bibinfo {volume} {6065},\ \bibinfo {editor} {edited by\ \bibinfo {editor}
  {\bibfnamefont {N.}~\bibnamefont {Adams}}, \bibinfo {editor} {\bibfnamefont
  {M.}~\bibnamefont {Berthold}}, \ and\ \bibinfo {editor} {\bibfnamefont
  {P.}~\bibnamefont {Cohen}}}\ (\bibinfo  {publisher} {Springer
  Berlin/Heidelberg},\ \bibinfo {year} {2010})\BibitemShut {NoStop}%
\bibitem [{\citenamefont {Mytkowicz}(2010)}]{todd-phd}%
  \BibitemOpen
  \bibfield  {author} {\bibinfo {author} {\bibfnamefont {T.}~\bibnamefont
  {Mytkowicz}},\ }\href@noop {} {Ph.D. thesis},\ \bibinfo  {school} {University
  of Colorado} (\bibinfo {year} {2010})\BibitemShut {NoStop}%
\bibitem [{\citenamefont {Henning}(2006)}]{spec}%
  \BibitemOpen
  \bibfield  {author} {\bibinfo {author} {\bibfnamefont {J.~L.}\ \bibnamefont
  {Henning}},\ }\href@noop {} {\bibfield  {journal} {\bibinfo  {journal}
  {SIGARCH Comput. Archit. News}\ }\textbf {\bibinfo {volume} {34}},\ \bibinfo
  {pages} {1} (\bibinfo {year} {2006})}\BibitemShut {NoStop}%
\bibitem [{\citenamefont {Anderson}\ \emph {et~al.}(1999)\citenamefont
  {Anderson}, \citenamefont {Bai}, \citenamefont {Bischof}, \citenamefont
  {Blackford}, \citenamefont {Demmel}, \citenamefont {Dongarra}, \citenamefont
  {Du~Croz}, \citenamefont {Greenbaum}, \citenamefont {Hammarling},
  \citenamefont {McKenney},\ and\ \citenamefont {Sorensen}}]{lapack}%
  \BibitemOpen
  \bibfield  {author} {\bibinfo {author} {\bibfnamefont {E.}~\bibnamefont
  {Anderson}}, \bibinfo {author} {\bibfnamefont {Z.}~\bibnamefont {Bai}},
  \bibinfo {author} {\bibfnamefont {C.}~\bibnamefont {Bischof}}, \bibinfo
  {author} {\bibfnamefont {S.}~\bibnamefont {Blackford}}, \bibinfo {author}
  {\bibfnamefont {J.}~\bibnamefont {Demmel}}, \bibinfo {author} {\bibfnamefont
  {J.}~\bibnamefont {Dongarra}}, \bibinfo {author} {\bibfnamefont
  {J.}~\bibnamefont {Du~Croz}}, \bibinfo {author} {\bibfnamefont
  {A.}~\bibnamefont {Greenbaum}}, \bibinfo {author} {\bibfnamefont
  {S.}~\bibnamefont {Hammarling}}, \bibinfo {author} {\bibfnamefont
  {A.}~\bibnamefont {McKenney}}, \ and\ \bibinfo {author} {\bibfnamefont
  {D.}~\bibnamefont {Sorensen}},\ }\href@noop {} {\emph {\bibinfo {title}
  {{LAPACK} Users' Guide}}},\ \bibinfo {edition} {3rd}\ ed.\ (\bibinfo
  {publisher} {Society for Industrial and Applied Mathematics},\ \bibinfo
  {address} {Philadelphia, PA},\ \bibinfo {year} {1999})\BibitemShut {NoStop}%
\bibitem [{\citenamefont {Hyndman}\ and\ \citenamefont
  {Khandakar}(2008)}]{autoARIMA}%
  \BibitemOpen
  \bibfield  {author} {\bibinfo {author} {\bibfnamefont {R.~J.}\ \bibnamefont
  {Hyndman}}\ and\ \bibinfo {author} {\bibfnamefont {Y.}~\bibnamefont
  {Khandakar}},\ }\href@noop {} {\bibfield  {journal} {\bibinfo  {journal} {J.
  Stat. Softw.}\ }\textbf {\bibinfo {volume} {27}},\ \bibinfo {pages} {1}
  (\bibinfo {year} {2008})}\BibitemShut {NoStop}%
\bibitem [{\citenamefont {Meese}\ and\ \citenamefont {Rogoff}(1983)}]{rwMeese}%
  \BibitemOpen
  \bibfield  {author} {\bibinfo {author} {\bibfnamefont {R.}~\bibnamefont
  {Meese}}\ and\ \bibinfo {author} {\bibfnamefont {K.}~\bibnamefont {Rogoff}},\
  }\href@noop {} {\bibfield  {journal} {\bibinfo  {journal} {J. Int. Econ.}\
  }\textbf {\bibinfo {volume} {14}},\ \bibinfo {pages} {3} (\bibinfo {year}
  {1983})}\BibitemShut {NoStop}%
\bibitem [{\citenamefont {Mu{\'c}k}\ and\ \citenamefont
  {Skrzypczy{\'n}ski}(2012)}]{rwCCE}%
  \BibitemOpen
  \bibfield  {author} {\bibinfo {author} {\bibfnamefont {J.}~\bibnamefont
  {Mu{\'c}k}}\ and\ \bibinfo {author} {\bibfnamefont {P.}~\bibnamefont
  {Skrzypczy{\'n}ski}},\ }\href@noop {} {}\bibinfo {type} {National Bank of
  Poland Working Papers}\ \bibinfo {number} {127}\ (\bibinfo {year} {2012})\
  \bibinfo {note} {(unpublished)}\BibitemShut {NoStop}%
\bibitem [{\citenamefont {Brockwell}\ and\ \citenamefont
  {Davis}(2002)}]{davislinearts}%
  \BibitemOpen
  \bibfield  {author} {\bibinfo {author} {\bibfnamefont {P.~J.}\ \bibnamefont
  {Brockwell}}\ and\ \bibinfo {author} {\bibfnamefont {R.~A.}\ \bibnamefont
  {Davis}},\ }\href@noop {} {\emph {\bibinfo {title} {Introduction to Time
  Series and Forecasting}}}\ (\bibinfo  {publisher} {Springer-Verlag},\
  \bibinfo {year} {2002})\BibitemShut {NoStop}%
\bibitem [{\citenamefont {Kwiatkowski}\ \emph {et~al.}(1992)\citenamefont
  {Kwiatkowski}, \citenamefont {Phillips}, \citenamefont {Schmidt},\ and\
  \citenamefont {Shin}}]{KPSSunit}%
  \BibitemOpen
  \bibfield  {author} {\bibinfo {author} {\bibfnamefont {D.}~\bibnamefont
  {Kwiatkowski}}, \bibinfo {author} {\bibfnamefont {P.~C.~B.}\ \bibnamefont
  {Phillips}}, \bibinfo {author} {\bibfnamefont {P.}~\bibnamefont {Schmidt}}, \
  and\ \bibinfo {author} {\bibfnamefont {Y.}~\bibnamefont {Shin}},\ }\href@noop
  {} {\bibfield  {journal} {\bibinfo  {journal} {J. Econometrics}\ }\textbf
  {\bibinfo {volume} {54}},\ \bibinfo {pages} {159} (\bibinfo {year}
  {1992})}\BibitemShut {NoStop}%
\bibitem [{\citenamefont {Canova}\ and\ \citenamefont
  {Hansen}(1995)}]{Canova1995}%
  \BibitemOpen
  \bibfield  {author} {\bibinfo {author} {\bibfnamefont {F.}~\bibnamefont
  {Canova}}\ and\ \bibinfo {author} {\bibfnamefont {B.~E.}\ \bibnamefont
  {Hansen}},\ }\href@noop {} {\bibfield  {journal} {\bibinfo  {journal} {J.
  Bus. Econ. Stat.}\ }\textbf {\bibinfo {volume} {13}},\ \bibinfo {pages} {237}
  (\bibinfo {year} {1995})}\BibitemShut {NoStop}%
\bibitem [{\citenamefont {Akaike}(1974)}]{akaike1974}%
  \BibitemOpen
  \bibfield  {author} {\bibinfo {author} {\bibfnamefont {H.}~\bibnamefont
  {Akaike}},\ }\href@noop {} {\bibfield  {journal} {\bibinfo  {journal} {{IEEE}
  Trans. Automat. Contr.}\ }\textbf {\bibinfo {volume} {19}},\ \bibinfo {pages}
  {716} (\bibinfo {year} {1974})}\BibitemShut {NoStop}%
\bibitem [{\citenamefont {Packard}\ \emph {et~al.}(1980)\citenamefont
  {Packard}, \citenamefont {Crutchfield}, \citenamefont {Farmer},\ and\
  \citenamefont {Shaw}}]{packard80}%
  \BibitemOpen
  \bibfield  {author} {\bibinfo {author} {\bibfnamefont {N.~H.}\ \bibnamefont
  {Packard}}, \bibinfo {author} {\bibfnamefont {J.~P.}\ \bibnamefont
  {Crutchfield}}, \bibinfo {author} {\bibfnamefont {J.~D.}\ \bibnamefont
  {Farmer}}, \ and\ \bibinfo {author} {\bibfnamefont {R.~S.}\ \bibnamefont
  {Shaw}},\ }\href@noop {} {\bibfield  {journal} {\bibinfo  {journal} {Phys.
  Rev. Lett.}\ }\textbf {\bibinfo {volume} {45}},\ \bibinfo {pages} {712}
  (\bibinfo {year} {1980})}\BibitemShut {NoStop}%
\bibitem [{\citenamefont {Sauer}\ \emph {et~al.}(1991)\citenamefont {Sauer},
  \citenamefont {Yorke},\ and\ \citenamefont {Casdagli}}]{Sauer:1991lr}%
  \BibitemOpen
  \bibfield  {author} {\bibinfo {author} {\bibfnamefont {T.}~\bibnamefont
  {Sauer}}, \bibinfo {author} {\bibfnamefont {J.}~\bibnamefont {Yorke}}, \ and\
  \bibinfo {author} {\bibfnamefont {M.}~\bibnamefont {Casdagli}},\ }\href@noop
  {} {\bibfield  {journal} {\bibinfo  {journal} {J. Stat. Phys.}\ }\textbf
  {\bibinfo {volume} {65}},\ \bibinfo {pages} {579} (\bibinfo {year}
  {1991})}\BibitemShut {NoStop}%
\bibitem [{\citenamefont {Takens}(1981)}]{Takens:1981uq}%
  \BibitemOpen
  \bibfield  {author} {\bibinfo {author} {\bibfnamefont {F.}~\bibnamefont
  {Takens}},\ }in\ \href@noop {} {\emph {\bibinfo {booktitle} {Dynamical
  Systems and Turbulence, Warwick 1980}}},\ Vol.\ \bibinfo {volume} {898},\
  \bibinfo {editor} {edited by\ \bibinfo {editor} {\bibfnamefont
  {D.}~\bibnamefont {Rand}}\ and\ \bibinfo {editor} {\bibfnamefont {L.-S.}\
  \bibnamefont {Young}}}\ (\bibinfo  {publisher} {Springer Berlin/Heidelberg},\
  \bibinfo {year} {1981})\ pp.\ \bibinfo {pages} {366--381}\BibitemShut
  {NoStop}%
\bibitem [{\citenamefont {Fraser}\ and\ \citenamefont
  {Swinney}(1986)}]{fraser-swinney}%
  \BibitemOpen
  \bibfield  {author} {\bibinfo {author} {\bibfnamefont {A.~M.}\ \bibnamefont
  {Fraser}}\ and\ \bibinfo {author} {\bibfnamefont {H.~L.}\ \bibnamefont
  {Swinney}},\ }\href@noop {} {\bibfield  {journal} {\bibinfo  {journal} {Phys.
  Rev. A}\ }\textbf {\bibinfo {volume} {33}},\ \bibinfo {pages} {1134}
  (\bibinfo {year} {1986})}\BibitemShut {NoStop}%
\bibitem [{\citenamefont {Kennel}\ \emph {et~al.}(1992)\citenamefont {Kennel},
  \citenamefont {Brown},\ and\ \citenamefont {Abarbanel}}]{KBA92}%
  \BibitemOpen
  \bibfield  {author} {\bibinfo {author} {\bibfnamefont {M.~B.}\ \bibnamefont
  {Kennel}}, \bibinfo {author} {\bibfnamefont {R.}~\bibnamefont {Brown}}, \
  and\ \bibinfo {author} {\bibfnamefont {H.~D.~I.}\ \bibnamefont {Abarbanel}},\
  }\href@noop {} {\bibfield  {journal} {\bibinfo  {journal} {Phys. Rev. A}\
  }\textbf {\bibinfo {volume} {45}},\ \bibinfo {pages} {3403} (\bibinfo {year}
  {1992})}\BibitemShut {NoStop}%
\bibitem [{\citenamefont {Weigend}\ and\ \citenamefont
  {Gershenfeld}(1993)}]{weigend-book}%
  \BibitemOpen
  \bibinfo {editor} {\bibfnamefont {A.}~\bibnamefont {Weigend}}\ and\ \bibinfo
  {editor} {\bibfnamefont {N.}~\bibnamefont {Gershenfeld}},\ eds.,\ \href@noop
  {} {\emph {\bibinfo {title} {Time Series Prediction: Forecasting the Future
  and Understanding the Past}}}\ (\bibinfo  {publisher} {Santa Fe Institute},\
  \bibinfo {year} {1993})\BibitemShut {NoStop}%
\bibitem [{\citenamefont {Casdagli}\ and\ \citenamefont
  {Eubank}(1992)}]{casdagli-eubank92}%
  \BibitemOpen
  \bibinfo {editor} {\bibfnamefont {M.}~\bibnamefont {Casdagli}}\ and\ \bibinfo
  {editor} {\bibfnamefont {S.}~\bibnamefont {Eubank}},\ eds.,\ \href@noop {}
  {\emph {\bibinfo {title} {Nonlinear Modeling and Forecasting}}}\ (\bibinfo
  {publisher} {Addison Wesley},\ \bibinfo {year} {1992})\BibitemShut {NoStop}%
\bibitem [{\citenamefont {Pikovsky}(1986)}]{pikovsky86-sov}%
  \BibitemOpen
  \bibfield  {author} {\bibinfo {author} {\bibfnamefont {A.}~\bibnamefont
  {Pikovsky}},\ }\href@noop {} {\bibfield  {journal} {\bibinfo  {journal} {Sov.
  J. Commun. Technol. Electron.}\ }\textbf {\bibinfo {volume} {31}},\ \bibinfo
  {pages} {911} (\bibinfo {year} {1986})}\BibitemShut {NoStop}%
\bibitem [{\citenamefont {Sugihara}\ and\ \citenamefont
  {May}(1990)}]{sugihara90}%
  \BibitemOpen
  \bibfield  {author} {\bibinfo {author} {\bibfnamefont {G.}~\bibnamefont
  {Sugihara}}\ and\ \bibinfo {author} {\bibfnamefont {R.~M.}\ \bibnamefont
  {May}},\ }\href@noop {} {\bibfield  {journal} {\bibinfo  {journal} {Nature}\
  }\textbf {\bibinfo {volume} {344}},\ \bibinfo {pages} {734} (\bibinfo {year}
  {1990})}\BibitemShut {NoStop}%
\bibitem [{\citenamefont {Lorenz}(1969)}]{lorenz-analogues}%
  \BibitemOpen
  \bibfield  {author} {\bibinfo {author} {\bibfnamefont {E.~N.}\ \bibnamefont
  {Lorenz}},\ }\href@noop {} {\bibfield  {journal} {\bibinfo  {journal} {J.
  Atmos. Sci.}\ }\textbf {\bibinfo {volume} {26}},\ \bibinfo {pages} {636}
  (\bibinfo {year} {1969})}\BibitemShut {NoStop}%
\bibitem [{\citenamefont {Hyndman}\ and\ \citenamefont {Koehler}(2006)}]{MASE}%
  \BibitemOpen
  \bibfield  {author} {\bibinfo {author} {\bibfnamefont {R.~J.}\ \bibnamefont
  {Hyndman}}\ and\ \bibinfo {author} {\bibfnamefont {A.~B.}\ \bibnamefont
  {Koehler}},\ }\href@noop {} {\bibfield  {journal} {\bibinfo  {journal} {Int.
  J. Forecasting}\ }\textbf {\bibinfo {volume} {22}},\ \bibinfo {pages} {679}
  (\bibinfo {year} {2006})}\BibitemShut {NoStop}%
\bibitem [{\citenamefont {Pesin}(1977)}]{pesin77}%
  \BibitemOpen
  \bibfield  {author} {\bibinfo {author} {\bibfnamefont {Y.~B.}\ \bibnamefont
  {Pesin}},\ }\href@noop {} {\bibfield  {journal} {\bibinfo  {journal} {Russ.
  Math. Surv.}\ }\textbf {\bibinfo {volume} {32}},\ \bibinfo {pages} {55}
  (\bibinfo {year} {1977})}\BibitemShut {NoStop}%
\bibitem [{\citenamefont {Petersen}(1989)}]{petersen1989}%
  \BibitemOpen
  \bibfield  {author} {\bibinfo {author} {\bibfnamefont {K.}~\bibnamefont
  {Petersen}},\ }\href@noop {} {\emph {\bibinfo {title} {Ergodic Theory}}}\
  (\bibinfo  {publisher} {Cambridge University Press},\ \bibinfo {year}
  {1989})\BibitemShut {NoStop}%
\bibitem [{\citenamefont {Amig{\'o}}(2012)}]{amigo2012permutation}%
  \BibitemOpen
  \bibfield  {author} {\bibinfo {author} {\bibfnamefont {J.}~\bibnamefont
  {Amig{\'o}}},\ }\href@noop {} {\emph {\bibinfo {title} {Permutation
  Complexity in Dynamical Systems: Ordinal Patterns, Permutation Entropy and
  All That}}}\ (\bibinfo  {publisher} {Springer},\ \bibinfo {year}
  {2012})\BibitemShut {NoStop}%
\bibitem [{\citenamefont {Fadlallah}\ \emph {et~al.}(2013)\citenamefont
  {Fadlallah}, \citenamefont {Chen}, \citenamefont {Keil},\ and\ \citenamefont
  {Pr{\'\i}ncipe}}]{fadlallah2013}%
  \BibitemOpen
  \bibfield  {author} {\bibinfo {author} {\bibfnamefont {B.}~\bibnamefont
  {Fadlallah}}, \bibinfo {author} {\bibfnamefont {B.}~\bibnamefont {Chen}},
  \bibinfo {author} {\bibfnamefont {A.}~\bibnamefont {Keil}}, \ and\ \bibinfo
  {author} {\bibfnamefont {J.}~\bibnamefont {Pr{\'\i}ncipe}},\ }\href@noop {}
  {\bibfield  {journal} {\bibinfo  {journal} {Phys. Rev. E}\ }\textbf {\bibinfo
  {volume} {87}},\ \bibinfo {pages} {022911} (\bibinfo {year}
  {2013})}\BibitemShut {NoStop}%
\bibitem [{\citenamefont {H\'{e}non}(1976)}]{henon}%
  \BibitemOpen
  \bibfield  {author} {\bibinfo {author} {\bibfnamefont {M.}~\bibnamefont
  {H\'{e}non}},\ }\href@noop {} {\bibfield  {journal} {\bibinfo  {journal}
  {Communications in Mathematical Physics}\ }\textbf {\bibinfo {volume} {50}},\
  \bibinfo {pages} {69} (\bibinfo {year} {1976})}\BibitemShut {NoStop}%
\bibitem [{\citenamefont {Lorenz}(1963)}]{lorenz}%
  \BibitemOpen
  \bibfield  {author} {\bibinfo {author} {\bibfnamefont {E.}~\bibnamefont
  {Lorenz}},\ }\href@noop {} {\bibfield  {journal} {\bibinfo  {journal}
  {Journal of the Atmospheric Sciences}\ }\textbf {\bibinfo {volume} {20}},\
  \bibinfo {pages} {130} (\bibinfo {year} {1963})}\BibitemShut {NoStop}%
\bibitem [{\citenamefont {Cao}\ \emph {et~al.}(2004)\citenamefont {Cao},
  \citenamefont {Tung}, \citenamefont {Gao}, \citenamefont {Protopopescu},\
  and\ \citenamefont {Hively}}]{cao2004det}%
  \BibitemOpen
  \bibfield  {author} {\bibinfo {author} {\bibfnamefont {Y.}~\bibnamefont
  {Cao}}, \bibinfo {author} {\bibfnamefont {W.~W.}\ \bibnamefont {Tung}},
  \bibinfo {author} {\bibfnamefont {J.~B.}\ \bibnamefont {Gao}}, \bibinfo
  {author} {\bibfnamefont {V.~A.}\ \bibnamefont {Protopopescu}}, \ and\
  \bibinfo {author} {\bibfnamefont {L.~M.}\ \bibnamefont {Hively}},\
  }\href@noop {} {\bibfield  {journal} {\bibinfo  {journal} {Phys. Rev. E}\
  }\textbf {\bibinfo {volume} {70}},\ \bibinfo {pages} {046217} (\bibinfo
  {year} {2004})}\BibitemShut {NoStop}%
\bibitem [{\citenamefont {Goebel}\ \emph {et~al.}(2009)\citenamefont {Goebel},
  \citenamefont {Sanfelice},\ and\ \citenamefont {Teel}}]{hybrid}%
  \BibitemOpen
  \bibfield  {author} {\bibinfo {author} {\bibfnamefont {R.}~\bibnamefont
  {Goebel}}, \bibinfo {author} {\bibfnamefont {R.}~\bibnamefont {Sanfelice}}, \
  and\ \bibinfo {author} {\bibfnamefont {A.}~\bibnamefont {Teel}},\ }\href@noop
  {} {\bibfield  {journal} {\bibinfo  {journal} {{IEEE} Control Syst. Mag.}\
  }\textbf {\bibinfo {volume} {29}},\ \bibinfo {pages} {28} (\bibinfo {year}
  {2009})}\BibitemShut {NoStop}%
\end{thebibliography}%

\end{document}